\documentclass[%
    aps, 
    prab, 
    twocolumn, 
    superscriptaddress, 
    longbibliography,
    floatfix,
    10pt
]{revtex4-2}

\usepackage{graphicx}
\usepackage{dcolumn}
\usepackage{bm}
\usepackage{overpic}
\usepackage{relsize}
\usepackage{amsmath}
\usepackage{amssymb}

\begin{document}

\preprint{APS/123-QED}

\title{Towards precision quantitative measurement of radiation reaction within the classical radiation-dominated regime  }

\author{Minghao Ma}
\thanks{These authors have contributed equally to this work.}
\affiliation{State Key Laboratory of Dark Matter Physics, Key Laboratory for Laser Plasmas (MoE), School of Physics and Astronomy, Shanghai Jiao Tong University, Shanghai 200240, China}

\author{Ke Liu}
\thanks{These authors have contributed equally to this work.}
\affiliation{College of Science, National University of Defense Technology, Changsha 410073, China}

\author{Ge Zhou}
\affiliation{State Key Laboratory of Dark Matter Physics, Key Laboratory for Laser Plasmas (MoE), School of Physics and Astronomy, Shanghai Jiao Tong University, Shanghai 200240, China}

\author{Zhida Yang}
\affiliation{State Key Laboratory of Dark Matter Physics, Key Laboratory for Laser Plasmas (MoE), School of Physics and Astronomy, Shanghai Jiao Tong University, Shanghai 200240, China}

\author{Yulin Xin}
\affiliation{College of Science, National University of Defense Technology, Changsha 410073, China}

\author{Jiadong Yang}
\affiliation{College of Science, National University of Defense Technology, Changsha 410073, China}

\author{Pengfei Zhu}
\affiliation{State Key Laboratory of Dark Matter Physics, Key Laboratory for Laser Plasmas (MoE), School of Physics and Astronomy, Shanghai Jiao Tong University, Shanghai 200240, China}
\affiliation{Tsung-Dao Lee Institute, Shanghai Jiao Tong University, Shanghai 201210, China}

\author{Yipeng Wu}
\affiliation{State Key Laboratory of Dark Matter Physics, Key Laboratory for Laser Plasmas (MoE), School of Physics and Astronomy, Shanghai Jiao Tong University, Shanghai 200240, China}
\affiliation{Tsung-Dao Lee Institute, Shanghai Jiao Tong University, Shanghai 201210, China}

\author{Min Chen}
\affiliation{State Key Laboratory of Dark Matter Physics, Key Laboratory for Laser Plasmas (MoE), School of Physics and Astronomy, Shanghai Jiao Tong University, Shanghai 200240, China}
\affiliation{Collaborative Innovation Center of IFSA, Shanghai Jiao Tong University, Shanghai 200240, China
}

\author{Tongpu Yu}
\email{tongpu@nudt.edu.cn}
\affiliation{College of Science, National University of Defense Technology, Changsha 410073, China}

\author{Wenchao Yan}
\email{wenchaoyan@sjtu.edu.cn}
\affiliation{State Key Laboratory of Dark Matter Physics, Key Laboratory for Laser Plasmas (MoE), School of Physics and Astronomy, Shanghai Jiao Tong University, Shanghai 200240, China}
\affiliation{Collaborative Innovation Center of IFSA, Shanghai Jiao Tong University, Shanghai 200240, China
}

\author{Jie Zhang}
\email{jzhang1@sjtu.edu.cn}
\affiliation{State Key Laboratory of Dark Matter Physics, Key Laboratory for Laser Plasmas (MoE), School of Physics and Astronomy, Shanghai Jiao Tong University, Shanghai 200240, China}
\affiliation{Tsung-Dao Lee Institute, Shanghai Jiao Tong University, Shanghai 201210, China}
\affiliation{Collaborative Innovation Center of IFSA, Shanghai Jiao Tong University, Shanghai 200240, China
}

\date{\today}

\begin{abstract}
Radiation reaction (RR) is a fundamental yet incompletely validated process in laser-particle interactions, since it lacks quantitatively definitive experimental verifications, especially the transition from classical to quantum regime. Herein, we propose a novel experimental scenario for investigating radiation RR within the classical radiation-dominated regime (CRDR), via the collision of a high-intensity petawatt-class laser with a tens-of-MeV electron beam from a LINAC. This approach enables access to a distinct parameter regime wherein RR dominates electron dynamics while quantum effects remain modest. Numerical simulations demonstrate that three key observables exist for identifying the RR within this CRDR regime: (i) quantitative measurement of energy spectra to validate the quantum correction factor;  (ii) control of the collision time delay with charge-counting to map intensity dependence of RR; and (iii) verification of large angle ($90^\circ$) photon emission under the recoil condition $2\gamma \gtrsim a_0$. These experimental measurements will establish the benchmarks for RR models spanning the classical-to-quantum regime, thereby providing critical insights into fundamental strong-field quantum electrodynamics.

\end{abstract}

\maketitle


\section{introduction}

With the advancement of chirped pulse amplification (CPA) technology \cite{STRICKLAND1985219}, the laser power has reached beyond 10 petawatts or even hundreds of petawatts in the near future \cite{Danson2019HPLSE, Li2023LPR}. Laser facilities such as ELI \cite{eli}, SULF \cite{Li2018ol}, SILEX-II \cite{Hong.mre.2021}, Vulcan \cite{Danson.nf.2004,vulcan.2010}, and CORELS \cite{Sung2016IPC} are capable of reaching intensities exceeding $10^{22}\ \rm{W}/\rm{cm}^2$ or even $10^{23}\ \rm{W}/\rm{cm}^2$ \cite{dansonPetawattExawattClass2019a}.
These high-power lasers have significantly enhanced the intensity of particle-laser interaction, enabling important breakthroughs in experiments of laser-plasma interaction \cite{Picksley2024PRL, Chen2025HPLSE, Winkler2025Nature, Li.HPL.2025} and strong-field quantum electrodynamics (SFQED) \cite {Yan2017NP, Mirzaie2024NP, Cole2018PRX, Poder2018PRX, los.arXiv2407.12071}. The interaction between charged particles and intense electromagnetic fields has long been a cornerstone of classical and quantum electrodynamics \cite{Jackson.1998, Peskin.qft.1995}.  
First, charged particles will be violently accelerated in intense electromagnetic fields and emit electromagnetic radiations of wide spectrum.
These emissions then modify the dynamics of the charged particles due to the conservation of energy and momentum. 
This effect is well known as radiation reaction (RR) or radiation damping, which has been extensively investigated both theoretically and experimentally.

The RR effect was first systematically studied by Lorentz \cite{Lorentz1909}, Abraham \cite{Abraham1905} and Dirac \cite{Dirac1938} in the early 20th century.
The Lorentz-Abraham-Dirac (LAD) equation emerged as the relativistic generalization form of the charged particle motion equation with RR as an additional force beyond the Lorentz force.
Unfortunately, the LAD equation suffers from some well-known fundamental pathologies, including runaway solutions that predict unphysical exponential acceleration to infinite energy and pre-acceleration effects that violate causality by suggesting responses before forces are applied \cite{Ford1991PLA, blackburn2020radiation}.
These theoretical shortcomings motivated the search for alternative equations of motion with RR incorporated that could provide physically meaningful predictions while maintaining mathematical consistency \cite{Sokolov2009jetp,Burton2014cp}.
Landau and Lifshitz developed an artful approach by treating the radiation reaction as a perturbation to the Lorentz force \cite{Landau1975}. 
Their formulation eliminated the unphysical runaway solutions and pre-acceleration effects as mentioned above while maintaining agreement with the LAD equation in most physical scenarios. 
This is the Landau-Lifshitz (LL) equation, which has become the standard classical framework for modeling the RR effect, finding applications across diverse fields from fundamental physics to astrophysical plasmas \cite{petri2022AA, Tavani2011science, Jaroschek2009PRL} and laser-particle accelerators \cite{zhu2020sciadv, Duclous2011PPCF, Golovanov2022NJP, Ji2014PRL}.

The RR effect is inseparable from the radiation of electrons interacting with the electromagnetic field. 
When the interaction is weak, the energy loss caused by RR is usually negligible. Therefore, a high-interaction intensity is required to produce significant electron energy loss. 
In its average rest frame, an electron in an intense circularly polarized electromagnetic wave will be accelerated undergoing a circular motion \cite{Blackburn.rmpp.2020,Liu.epjd.2020}.
The fraction of energy radiated per cycle ($E_{rad}$) in relation to the electron energy in the average rest frame ($E_{e}$) can be used to quantify the strength of RR.
In the framework of classical electrodynamics \cite{Piazza.2008.lmp,Blackburn.rmpp.2020}, the invariant classical RR parameter is defined as 
\begin{eqnarray}
R_c = \frac{3E_{rad}}{4\pi E_{e}} = \alpha \eta_0 a^2_0.
\end{eqnarray}
Here, $\alpha \approx 1/137$ is the fine-structure constant, $a_0=eE/m_e\omega_0$ is the normalized field strength, $e$ is the elementary charge, $m_e$ is the electron mass, $\omega_0$ is the angular frequency of laser, and $\eta_0 = p_0 \cdot k_0 / m^2$ ($\hbar=c=1$) is the classical nonlinearity parameter, with the four-momentum $p_0$ and the wave four-vector $k_0$.  
To evaluate the significance of quantum radiation in the interaction, the quantum parameter is defined as \begin{eqnarray}
	\chi_{e} & = & \frac{\sqrt{-\left(F_{\mu\nu}p^{\nu} \right)^{2} }}{m_ecE_{cr}} \nonumber \\
	& = & \frac{\gamma}{E_{cr}}\sqrt{\left({\bf E} + {\bf v}\times{\bf B} \right)^{2} - \left({\bf E}\cdot{\bf v}/c \right)^{2}  }, \label{eqkaie}
\end{eqnarray}
where $F_{\mu\nu}$ is the EM field tensor, $p^{\nu}$ is the electron four-momentum, $E_{cr}={m_e^{2}c^{3}}/{e\hbar}$ is the Schwinger field\cite{PhysRev.82.664}. The Schwinger field $E_{cr}$ represents the field strength at which an electron can gain energy $m_ec^2$ within one Compton wavelength $\lambda_C=\hbar/m_{e}c$. Physically, $\chi_{e}$ represents the ratio of the field strength observed in the electron's instantaneous rest frame to $E_{cr}$.
When $R_c\sim1$ and $\chi_e\ll1$, the RR force is sufficiently strong to become comparable to the Lorentz force and significantly affect electron motion, yet quantum effects remain negligible. This condition defines the so-called classical radiation-dominated regime (CRDR) \cite{Piazza.rmp.2012}. 
When $\chi_e \sim 1$, the quantum effects become significant and the QED theory is necessitated\cite{Ritus.jrlr.1985,Gonoskov.rmp.2022,Yu2024rmpp, Geng2019CP}.
Unlike the continuous radiation process without an upper frequency limit in the classical framework,  quantum radiation theory introduces a high-frequency cutoff to ensure that the energy of a single emitted photon cannot exceed the parent electron energy \cite{Neitz2013PRL, Hu2020PRA, Blackburn.PhysRevLett.112.015001}.
Additionally, the quantum spin-flip transitions may further modify the radiation spectrum of photons.
Between classical and quantum theories, the semi-classical theory serves as a bridge by introducing a quantum correction factor $g(\chi_e)=P_{q}/P_{cl}$, where $P_{q}$ and $P_{cl}$ are the powers of quantum and classical radiation respectively \cite{ErberRevModPhys.38.626,Blackburn.rmpp.2020}. The factor $g(\chi_e)$ acts to reduce the overall radiation emission, thereby avoiding the problem where the radiated energy exceeds the electron energy within the classical framework.

\begin{figure*}
    \centering
    \includegraphics[width=1\textwidth]{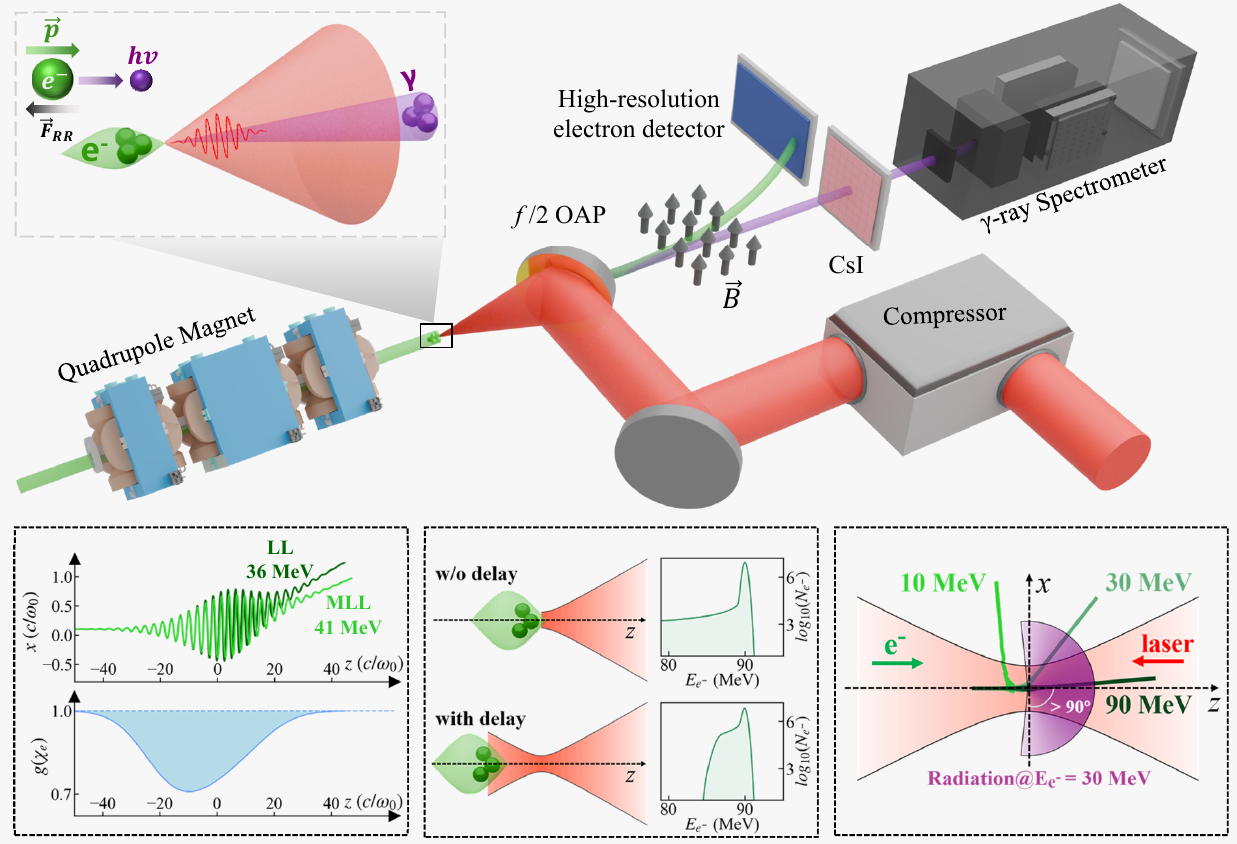}
    \put(-320,330){\larger{\larger{(a)}}}
    \put(-502,118){\larger{\larger{(b)}}}
    \put(-333,118){\larger{\larger{(c)}}}
    \put(-164,118){\larger{\larger{(d)}}}
    \caption{\label{fig:1} (a) Schematic of the proposed radiation reaction experiment at TDLI. The linear accelerator provides an electron bunch with a central energy of 10-90 MeV and an energy spread of 0.1\%, which collides with a 3 PW laser. The laser is focused by an Off-Axis Parabola (OAP) with an $f$-number of 2. After the collision, the electron energy spectrum is diagnosed by a high-precision magnetic spectrometer, while the $\gamma$-ray is diagnosed using CsI (Cesium Iodide) detectors and a spectrometer. (b) The single particle trajectories calculated based on the LL and MLL models respectively, as well as the envelope of correction factor $g(\chi_e)$ during the electron motion. (c) Adjusting the time delay between the electron bunch and the laser pulse can change the charge number participating in the reaction and the energy spectrum after the collision. (d) The trajectories of electrons with different initial energies $E_{e^-}$ after colliding with the laser, as well as the radiation angle for the electron bunch with $E_{e^-}=30\ \rm MeV$.}
\end{figure*}
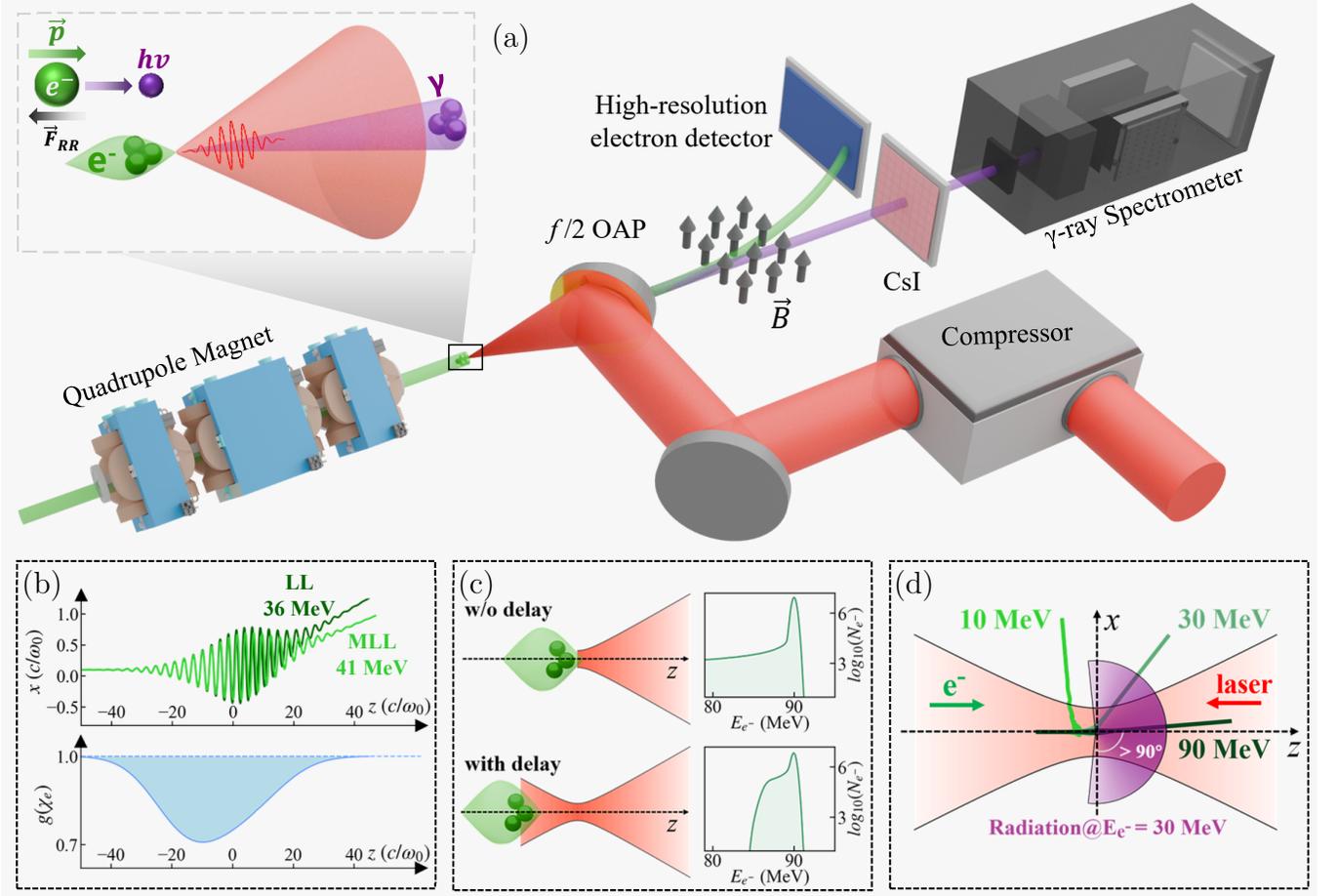

Experimentally testing of the accuracy of different RR models is one of the core issues in SFQED.
Especially, laser wakefield acceleration (LWFA) provides high-energy electron beams for the experimental verification of RR \cite{Tajima1979PRL, Vranic2014PRL, Samarin2017JMO}. To date, most reported experimental verifications of RR have relied on LWFA for high-energy electron beams \cite{Cole2018PRX, Poder2018PRX, los.arXiv2407.12071}. 
However, the laser-accelerated beams exhibit weaker stability, which poses challenges in determining the electron energy loss.
In contrast, electrons generated by traditional radio frequency accelerators exhibit high stability, flexible energy and high controllability. 
Note that the early E-144 experiments at SLAC observed nonlinear Compton scattering and pair production with 46.6 GeV electrons colliding with a $\sim 10^{18} \rm{W}/\rm{cm}^2$ laser, however, the RR effects were subdominant in these parameter ranges \cite{PhysRevLett.76.3116,PhysRevLett.79.1626,PhysRevD.60.092004}. 
Furthermore, several experimental schemes employing collisions between traditional accelerator electron beams and high-power lasers have been proposed to investigate SFQED, including the tests of different RR models.

The LUXE experiment at DESY is commissioned to utilize electron beams up to 16.5 GeV colliding with 350 TW lasers, with aims to achieve high quantum parameters ($\chi_e > 1$) \cite{Abramowicz2021, abramowicz2025inputesppuluxeexperiment}.
The E-320 collaboration of FACET-II at the SLAC National Accelerator Laboratory proposed SFQED experiments involving the collision of 10-13 GeV electron bunches with tightly focused 20 TW laser pulses, aiming to explore the quantum radiation reaction and other SFQED physics \cite{PhysRevLett.122.190404,PhysRevAccelBeams.22.101301,Chen:22, PhysRevLett.105.220403}.
These experiments primarily explore parameter regimes of high-energy electron beams and relatively moderate laser intensities, with the invariant quantum parameters $\chi_e \sim 1$ and the invariant classical radiation reaction parameter $R_c \sim 0.1$. However, experimental verification of strong RR within the classical framework where $\chi_e \sim 0.1$ but $R_c \sim 0.1$ is still lacking, which is crucial for the classical RR theory, as it represents the regime where RR is particularly strong before the onset of significant quantum corrections.


In the present paper, we introduce a unique RR experiment within the CRDR with $\chi_e \sim 0.1$ and $R_c \sim 0.1$, which is different from other SFQED processes dominated at higher $\chi_e$ values.
The combination of a high intensity PW laser ($\gtrsim 10^{22}\, \rm{W}/\rm{cm}^2$) and a moderate-energy linac electron beam (tens of MeV) enables intense laser-particle interactions but weak quantum effects.
The 3D particle-in-cell (PIC) simulations show that this parameter regime provides the access to the classical-quantum transition, making it ideal for testing semi-classical approaches compared with full classical and full QED predictions. 
Furthermore, variations in the parameters of the electron beam and laser pulse serve as direct evidence for verifying RR, including the generated radiation with a special angular distribution and the variations in the charge of low-energy electron beams after the collision.

\begin{figure*}
    \centering
    \includegraphics[width=1\textwidth]{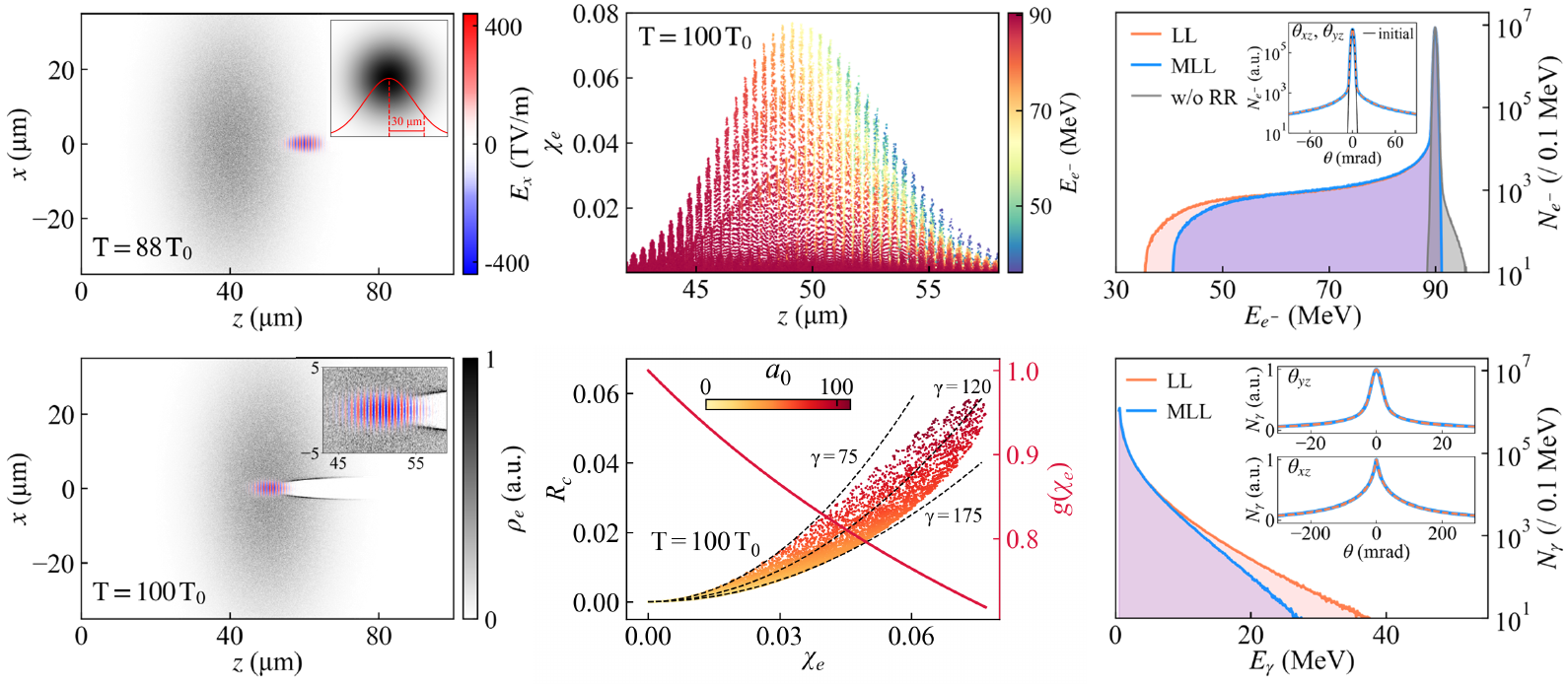}
    \put(-482,223){\larger{\larger{(a)}}}
    \put(-482,112){\larger{\larger{(b)}}}
    \put(-305,223){\larger{\larger{(c)}}}
    \put(-305,112){\larger{\larger{(d)}}}
    \put(-146,223){\larger{\larger{(e)}}}
    \put(-146,112){\larger{\larger{(f)}}}
    \caption{\label{fig:2} 3D-PIC simulation results of a 3 PW laser colliding with the 90 MeV linac electron bunch. (a-b) Snapshots of electron density distribution $\rho_e$ and the laser field $E_x$ in the polarized plane at time $t = 88\, \rm T_0$ and $t = 100\,\rm T_0$, respectively. The inset of (a) shows the initial transverse distribution of the electron bunch. The inset of (b) amplifies the laser field $E_x$ near the focus. (c) Distribution of $\chi_e$ at $t = 100\,\rm T_0$. The color bar represents the electron energy. Each point represents an individual electron. (d) Distribution of $R_c$ and $\chi_e$ for every electron at $t = 100\,\rm T_0$. The color bar represents the laser dimensionless parameter $a_0$ experienced by each electron. The black dashed lines show the variation of $R_c$ with $\chi_e$ when $\gamma = $  175, 120 and 75. The red line shows the Gaunt factor $g(\chi_e)$ at different $\chi_e$. (e) The electron energy spectrum after the collision. The gray line shows the spectrum without RR. The orange and blue lines represent the spectrum calculated based on the LL and MLL model, respectively. The inset shows the bunch divergence distribution before and after the collision. (f) The photon spectrum calculated using the two models. The inset shows the divergence angles of photons in the $yz$ plane (upper) and $xz$ plane (lower), respectively}
\end{figure*}

\section{method and models}
We start from the electron dynamics in laser fields.
Taking into account the interaction between an electron and the electromagnetic field with a four-vector RR force $g^{i}$, the equation of electron motion is governed by LL equation \cite{Landau1975}
\begin{eqnarray}
\frac{\mathrm{d}u^{i}}{\mathrm{d}s}&=&\frac{e}{mc^{2}}F^{ik}u_{k}+\frac{1}{mc}g^{i}, \label{eq:dui}\\
	g^{i}&=&\frac{2e^{3}}{3mc^{3}}\frac{\partial F^{ik}}{\partial x^{l}} u_{k}u^{l}-\frac{2e^{4}}{3m^{2}c^{5}}F^{il}F_{kl}u^{k} \nonumber \\
    && +\frac{2e^{4}}{3m^{2}c^{5}}\left(F_{kl}u^{l} \right) \left(F^{km}u_{m} \right)u^{i}, \label{llequation}
\end{eqnarray}
where $u_{i}=(\gamma, \textbf{p}/mc)$ is the electron four-velocity, $F^{ik}=\partial_{i}A_{k}-\partial_{k}A_{i}$ is the field tensor with $A_{i(k)}$ being the four-potential.
The semi-classical RR with quantum correction is described by the modified LL (MLL) equation. 
In this modified equation, the instantaneous power radiated by an electron is reduced by the factor $g(\chi_e)=P_{q}/P_{cl}$,
and the four-vector RR force $g^{i}$ is modified to $g(\chi_e)g^{i}$, where $g(\chi_e)$ is approximated by $g(\chi_e) \approx (3.7 \chi_e^3 + 31 \chi_e^2 + 12 \chi_e + 1)^{-4/9}$ for $\chi_e < 5$.\cite{Kirk2009PPCF, Thomas2012PRX}.

This new regime is currently accessible at the laboratory astrophysics experimental platform of Tsung-Dao Lee Institute (TDLI), which integrates a 3 PW ultra-intense femtosecond laser and a 90 MeV electron accelerator, thereby providing ideal conditions for experiments on radiation reaction within the CRDR. A schematic diagram of the proposed experimental setup is depicted in Fig.~\ref{fig:1}(a). A linearly polarized 800 nm laser pulse with a duration of 20 fs (Full Width at Half Maximum, FWHM) is focused by an Off-Axis Parabola (OAP) with an $f$-number of 2. The focused laser is considered to have an energy concentration of 50\%, and this configuration yields a peak intensity of $a_0=110$ with the beam waist radius $w_0=2\ \rm \mu m$. The linear accelerator has a center energy ranging flexibly from 10 to 90 MeV, a normalized emittance of 2 mm·mrad, and a longitudinal RMS energy spread $\sigma_{z}$ of 0.1\%. 

The primary observation relies on the distinct redistribution and broadening of the electron energy spectrum after the collision. Compared with the no scattering case, the distribution of electrons with energies lower than 90 MeV is expected to be absent, while this component appears in the energy spectrum when the colliding laser is present. This serves as a direct verification for RR. 
Diagnostic techniques for electron energy spectra and $\gamma$-ray energy spectra have gradually matured, which can provide a solid foundation for the feasibility of the experiment\cite{Du2024NP, Chen2022, lecoq2017inorganic, Corvan2014RSI, Hannasch2021SR, Schwinkendorf_2019}.
The three signatures for testing the RR models in the experiment are shown in fig.~\ref{fig:1}. Fig.~\ref{fig:1}(b) shows the final energy and trajectory of the electron in the Landau-Lifshitz (LL) and modified Landau-Lifshitz (MLL) models, respectively. According to the calculations, the maximum energy loss is 54 MeV with the LL model and 49 MeV with the MLL model. This $10\%$ difference and the correction factor $g(\chi_e)$ as low as 0.7 provides a clear and measurable discrimination between classical and semi-classical predictions. Fig.~\ref{fig:1}(c) demonstrates that direct control over the number of charges involved in the reaction can be achieved by adjusting the time delay between the electron bunch and the laser, enabling this parameter to be used for mapping the intensity dependence of RR. 
Fig. 1(d) illustrates that for electrons satisfying $2\gamma \gtrsim a_0$  (e.g., 30 MeV for $a_0\sim110$), radiation can be emitted at specific and resolved angles due to RR, thereby serving as a distinct and direct signature of the RR effect.

In the simulations, we use the fully relativistic EPOCH-3D \cite{RIDGERS2014273,Arber.ppcf.2025} with the simulation box of $35\ \rm \mu m $ (x) $\times\ 35\ \rm \mu m $ (y) $\times 100\ \rm \mu m $ (z), grid cells of $700 \times$ $700 \times$ $2000$, sampled by one macroparticle in each cell. The electron bunch moves in $z+$, whereas the laser is polarized in the $x$ direction and propagates along the $z-$ axis. The density distribution of the electron bunch can be approximated by 
\begin{equation}\label{eq:density}
\rho(x, y, z) = \rho_0 \exp\left(-\frac{x^2 + y^2}{w_e^2}\right) \exp\left(-\frac{(z/c)^2}{\tau_e^2}\right),
\end{equation}
where $w_e = 30\ \rm \mu m$, $\tau_e = 50\ \rm fs$ and $\rho_0$ is the peak density. The total charge of the bunch is 10 pC, and the center energy is 90 MeV. The longitudinal and transverse temperatures of electrons, $\epsilon_x=\epsilon_y=30 \ \rm keV$ and $\epsilon_z = 100 \ \rm keV$, are also considered to match the actual energy spread and emittance, ensuring the simulations faithfully represent the proposed experiment. In addition, the trajectory and the momentum of the single-electron are calculated based on Eq.~\ref{eq:dui} using a customized code, benchmarked by \cite{MZengponderomotive_scattergithub2022}. 
The tightly focused laser expression has been used in the code since the paraxial approximation breaks down near the focus for the tiny laser waist $w_0$ \cite{Salamin2002PRL}, where longitudinal field components would influence the electron dynamics.

\section{Results and discussion}


Figure~\ref{fig:2} presents the whole 3D-PIC simulation results of the collision between a 3 PW laser pulse and a 90 MeV electron bunch.
The snapshots of electron charge density distribution $\rho_e$ and the laser electric field $E_x$ in the polarization plane at $ t = 88\,\rm T_0$ and $ t = 100\,\rm T_0$ can be seen in Fig.~\ref{fig:2}(a,b). 
The electron beam collides head-on with the laser at the focus ($z=50\ \rm \mu m$), at $ t = 100\,\rm T_0$.
The distribution of $\chi_e$ and $R_c$ near the focus is shown in Fig.~\ref{fig:2}(c,d). Here, each point represents the result of a macro electron. 
When electrons encounter the focused PW laser, the radiation emission and its back-reaction on electrons occur simultaneously.
The periodic structure with a modulating envelope of $\chi_e$ along the $z$ direction Fig.~\ref{fig:2}(c) physically stems from the oscillating laser field and pulse envelope, which periodically affect the electron dynamics and thereby modulate the instantaneous emission rates.
Notably, the electrons experience significant energy depletion before encountering the peak laser intensity (evident from the color mapping in Fig.~\ref{fig:2}(c)).
Due to the small size of the laser focus, only a fraction of electrons experience the strongest laser field. This explains why most of the data points in Fig.~\ref{fig:2}(c) are clustered in the lower $\chi_e$ region ($\chi_e<0.01$). It is shown that during the whole interaction, the maximum $\chi_e$ can reach 0.077. 

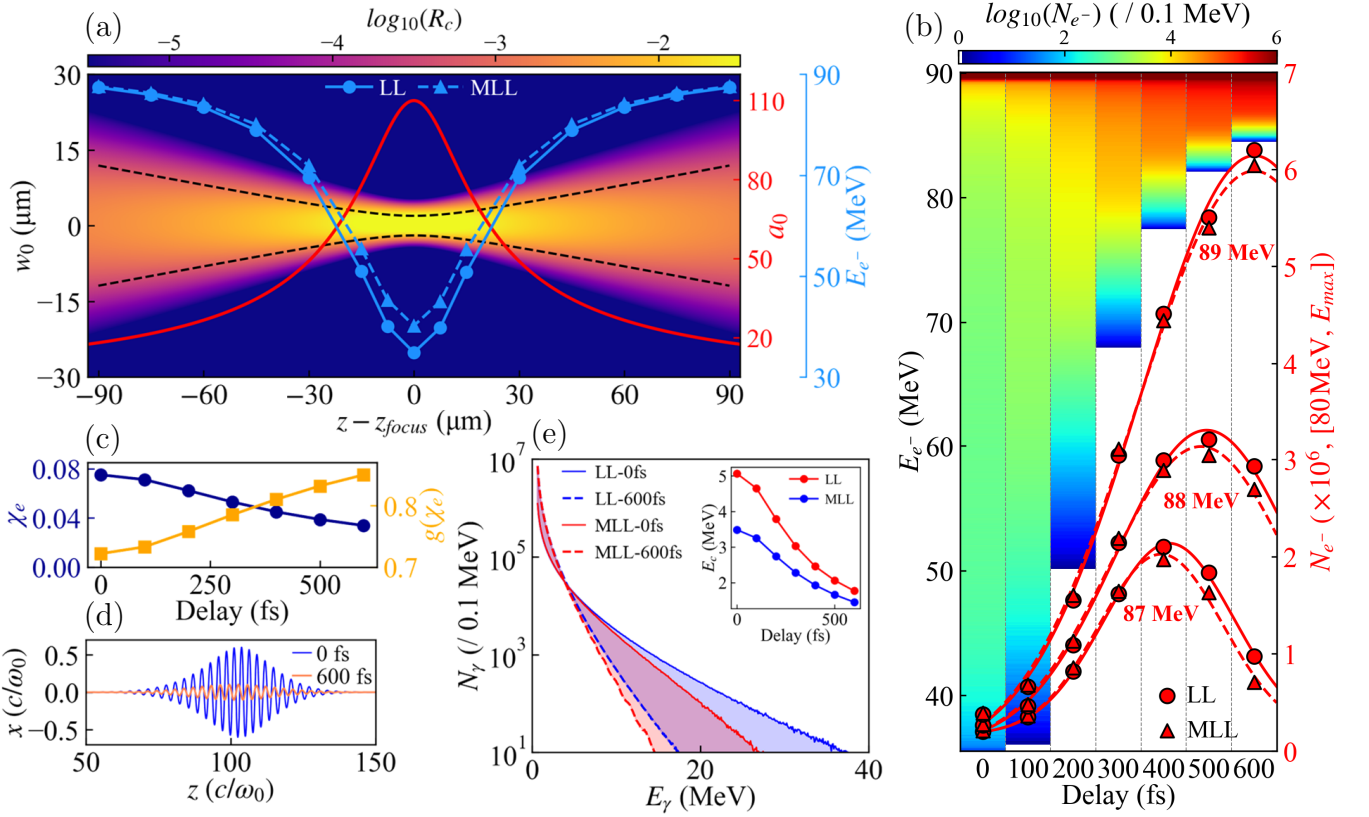
\begin{figure*}
    \centering
    \includegraphics[width=1\textwidth]{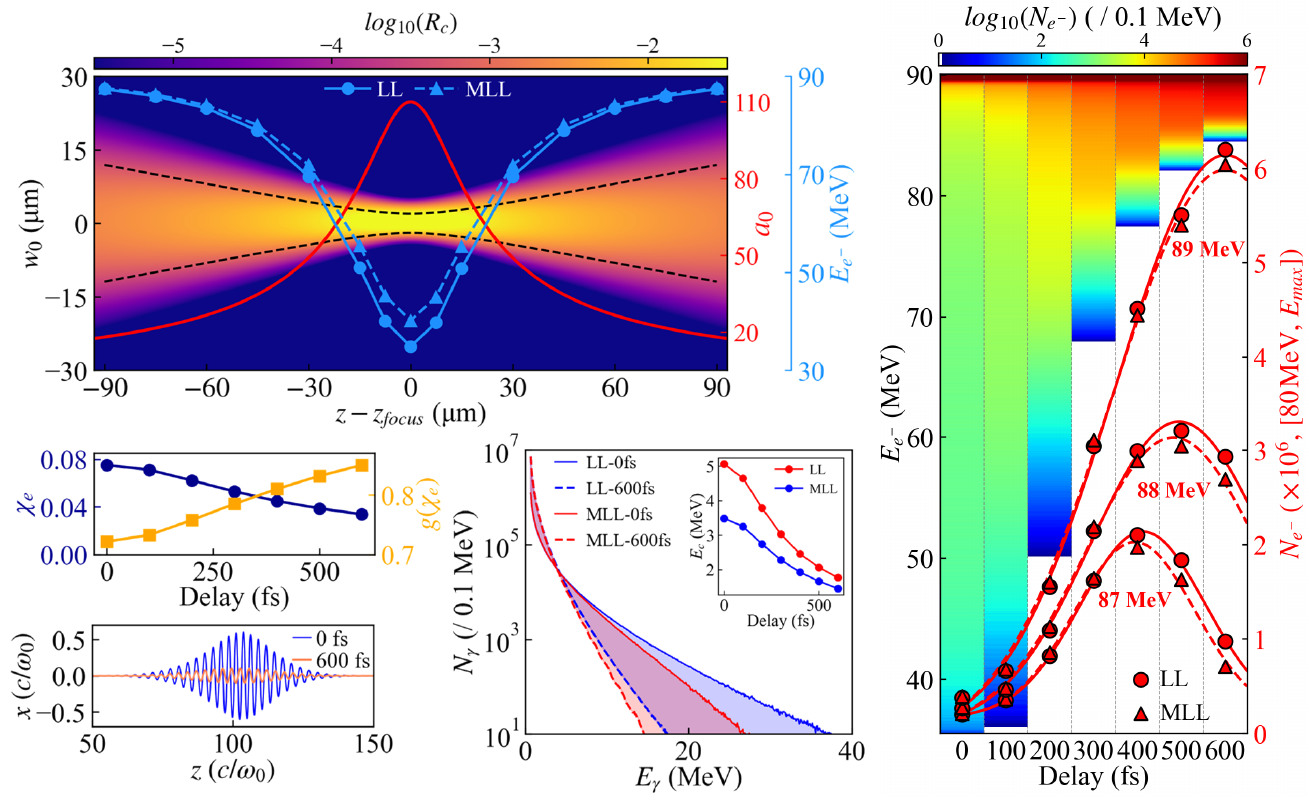}
    \put(-475,295){\larger{\larger{(a)}}}
    \put(-475,139){\larger{\larger{(c)}}}
    \put(-475,72){\larger{\larger{(d)}}}
    \put(-306,141){\larger{\larger{(e)}}}
    \put(-165,295){\larger{\larger{(b)}}}
    \caption{\label{fig:3} Simulation results by tuning the time delay between the laser pulse and the electron bunch. (a) The maximum $R_{c}$ in collisions of a laser at various defocusing positions $z-z_{focus}$ with a 90 MeV electron bunch. The black dashed lines shows the laser waist radius $w_0$ and the red line is the peak $a_0$ during the laser evolution. The blue solid and dashed lines show the final electron energy after the collision versus defocusing position of laser, as calculated by the LL and MLL models, respectively. (b) The electron spectrum after the collision at different time delays. The circular and triangular dots represent the total electron charge in different energy intervals. The red lines represent the Gaussian curve fitting. (c) The maximum $\chi_e$ and the minimal $g(\chi_e)$ during the interaction at different time delays. (d) The trajectory of an electron in head-on collision with laser. (e) The photon energy spectrum with a detection lower limit of 50 keV. The solid lines represent the spectrum without time delay, while the dashed lines represent the results with a 600 fs delay. The inset shows the critical photon energy versus the time delay.}
\end{figure*}

On the contrary, the parameter $R_c$ exhibits a distribution influenced by both the quantum parameter $\chi_e$ and the local laser parameter $a_0$. Overall, as $\chi_e$ increases, $R_c$ increases correspondingly, as shown in Fig.~\ref{fig:2}(d). 
Considering that $R_c \sim \alpha \chi_e a_0$ and $\chi_e \propto \gamma a_0$, when $\gamma$ remains constant, it can be concluded that $R_c \propto \chi_e^2$. The black dashed lines in Fig.~\ref{fig:2}(d) illustrate the variation of $R_c$ with $\chi_e$ when $\gamma$ is constant. With an initial $\gamma = 175$, the electrons begin to interact with the leading edge of the laser pulse, causing the energy to decrease. When $\gamma = 120$, the electrons are located at the focus of the laser, where both $R_c$ and $\chi_e$ reach the maximum values. After experiencing the full laser pulse, the electron energy drops to $\gamma = 75$. Since the electron energy changes continuously during the interaction process, the distribution of $R_c$ lies between the curves for $\gamma=175$ and $\gamma=75$.
The Gaunt factor $g(\chi_e)$ reaches a minimum of 0.7 at the focus (the red line in Fig.~\ref{fig:2}(d)), indicating a $30\%$ correction of the radiation reaction force and radiation power due to the quantum effect. 

The electron energy spectrum before and after the collision (shown in Fig.~\ref{fig:2}(e)) demonstrate a clear distinction between the LL and MLL models, and both cases differ significantly from the case without RR. In the absence of RR effects, electrons gain a slight energy increment through ponderomotive scattering without any energy loss. The maximal energy loss is consistent with the result of the single-electron trajectory calculations. Precise measurements of the electron energy spectrum below 50 MeV can be used to test the accuracy of semi-classical theory with $g(\chi_e)$.
The laser's ponderomotive force induces significant transverse scattering, resulting in a near-uniform angular distribution of electrons, as shown in the inset of Fig.~\ref{fig:2}(e).  

The photon energy spectrum is calculated with the lower detection limit of 50 keV in the simulation, as shown in Fig.~\ref{fig:2}(f). The classical RR predicts a higher cutoff photon energy of 38 MeV and greater high-energy photon yields compared to the semi-classical case of 28 MeV. These corrections in both the spectral cutoff and photon yield within the MLL model represent a direct impact of the quantum effect on nonlinear Thomson scattering, originating from the modified electron motion under a different RR force.
The inset of Fig.~\ref{fig:2}(f) illustrates the divergence angles of photons with a pronounced anisotropy characteristic. When $a_0 \gg 1$, the highly non-linear effects become prominent, leading to photon divergence angles of $\theta_{xz} \sim a_0/\gamma$ in the $xz$ plane significantly greater than $\theta_{yz} \sim 1/\gamma$ in the $yz$ plane \cite{Corde2013RMP}. 
This asymmetry occurs because although the photon emission is beamed into a $\sim 1/\gamma$ cone, the substantial transverse quiver motion in the polarization plane effectively broadens the emission angle along the $xy$ plane. This produces a unique non-circular radiation profile \cite{Yan2017NP}, which can serve as an additional clear experimental signature in experiments.

\begin{figure}
    \centering
    \includegraphics[width=0.48\textwidth]{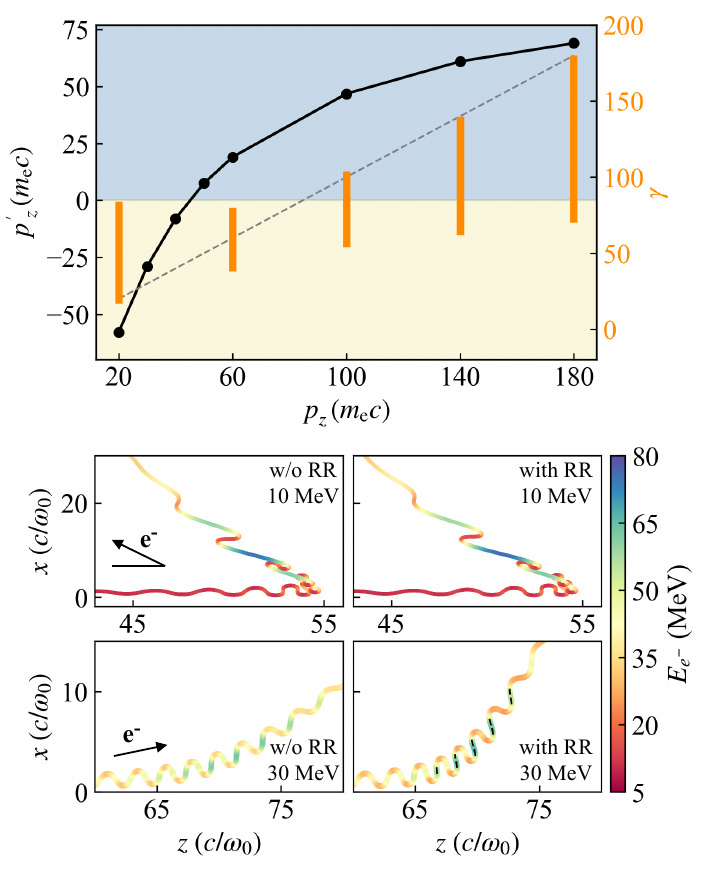}
    \put(-210,282){\larger{\larger{(a)}}}
    \put(-210,132){\larger{\larger{(b)}}}
    \put(-120,132){\larger{\larger{(c)}}}
    \put(-210,67){\larger{\larger{(d)}}}
    \put(-120,67){\larger{\larger{(e)}}}
    \caption{\label{fig:4} Simulation results with different initial longitudinal momentum $p_z$ for the LL model. (a) For different $p_z$, the black dots represent the minimal longitudinal momentum $p'_z$ of the electron bunch after the collision. The blue and  yellow regions are separated by $p'_z = 0$. The gray dashed line indicates $p_z = \gamma$, and the orange lines represent the range of electron energies after the collision. (b-e) The single electron's trajectory calculated via particle tracking code. The initial electron energy is 10 MeV in (b) and (c), and 30 MeV in (d) and (e). The black dashed lines in (e) represents the instantaneous $p_z < 0$.}
\end{figure}

Since the electron dynamics and the strength of RR are governed by the interaction intensity, adjusting this intensity allows for control over the observables.
Due to the tightly focusing property of the scattering laser, the spot expands rapidly and the intensity decreases after the focus. Therefore, introducing a precise time delay in the laser-electron collision enables control of the interaction intensity, which can directly affects the electron dynamics.
Fig. 3 presents the simulation results by tuning the time delay between the laser and the linac electron bunch. The black dashed line and the red line in fig.~\ref{fig:3}(a) show the waist and the peak electric field of the scattering laser, respectively. 
Colliding at the defocused position reduces the interaction intensity but increases the interaction region, allowing electrons far from the axis to experience a higher $R_c$, as shown in fig.~\ref{fig:3}(a). These centrally deviated 90 MeV electrons lose several MeV of energy after the collision. 
Fig.~\ref {fig:3}(b) presents the electron energy spectrum and charge counts in different energy intervals at varying time delays, where the charge number within a fixed energy interval depends significantly on the time delay. 
80 MeV is set as the lower threshold, and the upper detection limit is set as $E_{\text{max}} \leq 89\text{ MeV}$ to distinguish the initial energy spread of the electron bunch. 
A key observation is the distinct, non-monotonic behavior of the electron number in different energy intervals after the collision. When $E_{\text{max}} = 87$ and 88 MeV, the charge increases first and then decreases with longer delay; whereas at 89 MeV, it increases almost consistently. 
In addition, selecting a higher upper limit for the electron energy range causes a significant increase in the time delay corresponding to the peak number of charges. This trend can be understood by considering the enlarged laser spot at time delays, which engages more electrons from the periphery of the bunch into the RR-dominated interactions, thereby enhancing the detectable signal. While this comes at the cost of reducing the interaction intensity, it is beneficial for enhancing experimentally observable signals.

As the collision position deviates from the focus, the motion of electrons in the laser field differs significantly and directly affects the radiation energy. The energy of the electron colliding head-on with the laser has been calculated using the single-electron trajectory code as mentioned above. The blue line in Fig.~\ref{fig:3}(a) shows the electron energy after collision at different time delays for both the LL and the MLL model. The difference between the two cases is several MeV near the 0 fs time delay, while it becomes almost indistinguishable at higher delay times. This is consistent with the quantum correction decreasing with delay times, as demonstrated by the parameters $\chi_e$ and $g(\chi_e)$ in Fig.~\ref{fig:3}(c). 
The electron's trajectory in the $xz$ plane for the LL model at time delays of 0 and 600 fs is illuminated in Fig.~\ref{fig:3}(d). The electron exhibits a markedly larger amplitude of quiver motion at 0 time delay compared to the 600-fs case. This larger oscillation amplitude indicates stronger radiation emission, which is consistent with the higher local field strength at the laser focus. Fig.~\ref{fig:3}(e) illustrates the photon spectrum with the detection lower limit of 50 keV. The color-filled region represents the spectrum at time delays ranging from 0 to 600 fs. The progressive softening of the spectrum with increasing time delay is apparent. 
The critical photon energy $E_c$ for the radiation spectrum can be defined as \cite{Behm2018RSI}
\begin{equation}\label{eq:Ec}
\frac{dN_\gamma}{dE_\gamma} \propto E_\gamma^{-2/3} e^{-E_\gamma/E_{\text{c}}},
\end{equation}
which characterizes the exponential cutoff. 
The values of $E_c$  for the two models are basically equal at large time delays, as shown in the inset of Fig.~\ref{fig:3}(e).

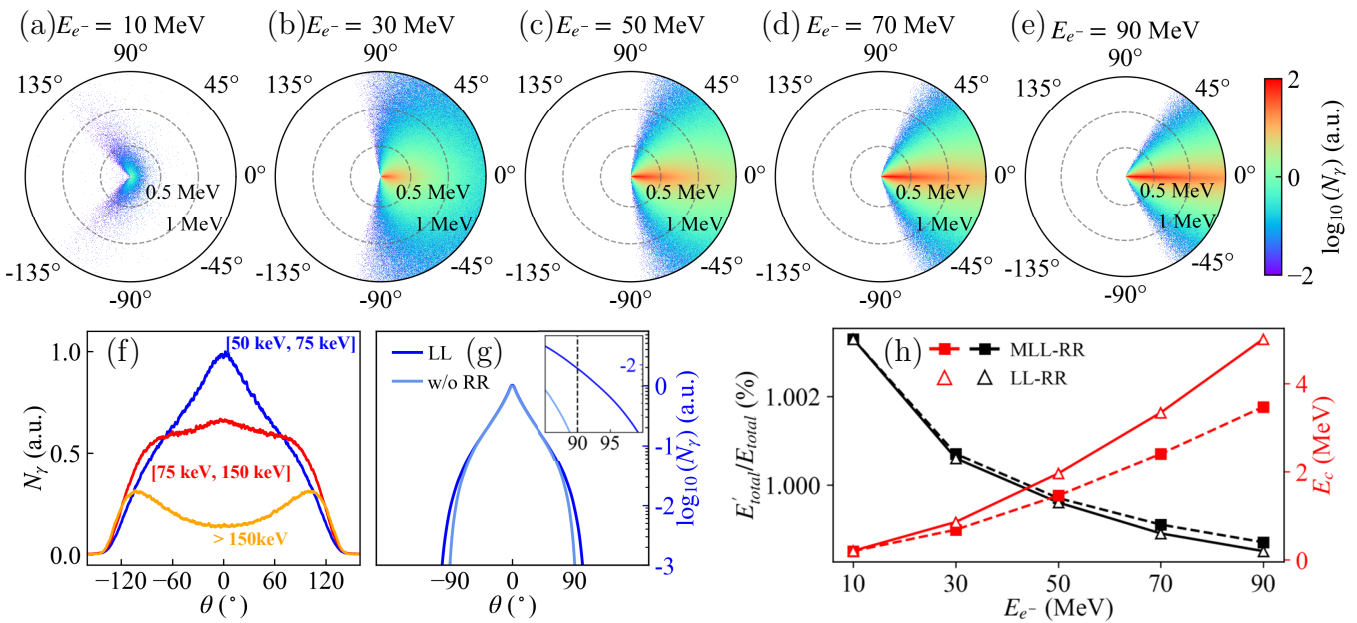
\begin{figure*}
    \centering
    \includegraphics[width=1\textwidth]{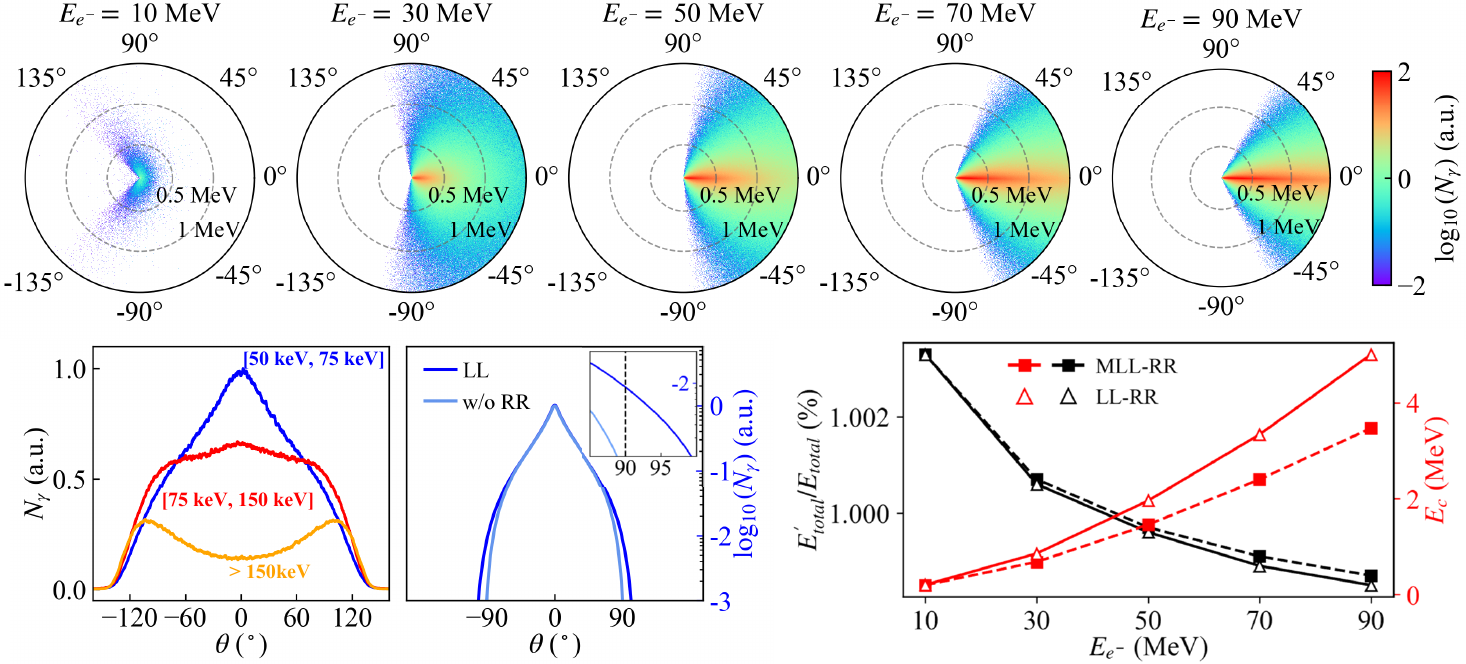}
    \put(-504,223){\larger{\larger{(a)}}}
    \put(-410,223){\larger{\larger{(b)}}}
    \put(-315,223){\larger{\larger{(c)}}}
    \put(-223,223){\larger{\larger{(d)}}}
    \put(-129,223){\larger{\larger{(e)}}}
    \put(-471,99){\larger{\larger{(f)}}}
    \put(-335,99){\larger{\larger{(g)}}}
    \put(-177,99){\larger{\larger{(h)}}}
    \caption{\label{fig:5} (a-e) The photon radiation results for different initial electron energies in the $xz$ plane. (f-g) The photon angle distribution for $E_{e^-} = 10\, \rm MeV$ (f) and $E_{e^-} = 30\, \rm MeV$ (g), based on the LL model. The inset in (g) shows the distribution around $90^{\circ}$, for cases with and without RR. (h) The black line shows the ratio of the minimum electron energy after collision to its initial value, and the critical photon energy is shown by the red line.}
\end{figure*}

Both the electron energy loss and the properties of the radiation tend to converge for the LL and MLL models with the increasing time delay in collision. This trend confirms that the reduced laser intensity at defocused positions suppresses the quantum effect, thereby reducing the difference between the two models. 
In addition, varying the time delay provides a method for detecting changes in charge and radiation, particularly several-fold differences in charge quantity and significant variations in critical photon energy $E_c$.
These detectable changes would also serve as a direct evidence of RR.

On the other hand, reducing the interaction intensity by decreasing the electron energy will significantly alter the electron trajectories, thereby affecting the radiation properties. It can be derived that when the electron energy is sufficiently low, the electron will change its direction of motion under the laser's ponderomotive force.
For the LL model, fig.~\ref{fig:4}(a) shows the bunch's minimum longitudinal momentum $p_z'$ after the collision with various initial longitudinal momentum $p_z$, without any time delay. 
The competition between RR and ponderomotive scattering depends critically on the electron's initial momentum. RR dominates momentum loss for higher $p_z$, whereas ponderomotive scattering dominates for lower $p_z$.

A key outcome of this analysis is the identification of a distinct electron recoil region, defined by $p_z'\leq0$ and highlighted by the light yellow area in fig.~\ref{fig:4}(a). 
The trajectory reveals a critical point at $p_z\sim \, 40m_ec$ and $p_z'\sim 0$, where a population of electrons comes to rest longitudinally after the collision. A further decrease in $p_z$ will cause $p_z'$ less than 0, implying that some electrons reverse their longitudinal motion direction after the collision. 
The orange stripe shows the energy range of the electron bunch after the collision. It can be seen that for the electrons with $p_z < 100\,m_ec$, the maximum final energy is higher than its initial value due, relative to the gray dashed line $p_z = \gamma$. This implies that the energy gained from the laser exceeds the energy loss due to RR.

To reveal the underlying mechanics, we use the single-particle trajectory code to explore the electron microscopic dynamics.
The electron is initially placed slightly off-axis before the collision. 
For a low-energy electron with $p_z = 20\,m_ec$ as shown in fig.~\ref{fig:4}(b-c), the interaction is dominated by the laser's ponderomotive force. The electron experiences drastic energy fluctuations, ultimately resulting in a reversed final longitudinal direction of motion. The electron trajectory is essentially consistent in both the presence and absence of RR, which also indicates that the RR effect is not the primary cause for the reversing of motion in this ponderomotive-dominated regime. 
In contrast, the dynamics for a higher-energy electron with $p_z = 60\,m_ec$ (fig.~\ref{fig:4}(d-e)) reveal a distinct signature of RR.
Although no final longitudinal motion reversal occurs and energy fluctuation decreases, the trajectory with RR (fig.~\ref{fig:4}(e)) exists a transient region of reverse longitudinal motion as indicated by the black dashed line, which is absent in the non-RR case (fig.~\ref{fig:4}(d)). 
This transient reversal, occurring under the condition that $2\gamma \gtrsim a_0$, can lead to the photon divergence angles exceeding $90^\circ$ in the polarization plane, which can be regarded as a strong signal of RR below the CRDR, as stated in Ref. \cite{Piazza2009PRL}.

The most definitive experimental signature of electron recoil is the emergence of large-angle photon emission.
Fig.~\ref{fig:5}(a-e) shows the evolution of photon distribution within the laser polarization $xz$ plane for the LL model. As the electron energy $E_{e^{-}}$ increases, the divergence angle of the emitted photons decreases while the total yield increases. 
The energy-resolved angular distribution offers deep insights into the underlying physics. 
Fig.~\ref{fig:5}(f) shows the distribution of photons across different energy intervals when the incident electron energy is 10 MeV. The high-energy photons population peaks around $\pm110^\circ$.
This backward emission is a direct result of electrons that have been accelerated in the reverse direction by the laser field as mentioned in Fig.~\ref{fig:4}(b-c).
In contrast, low-energy photons are concentrated around $0^\circ$, corresponding to forward emission from the early interaction before significant recoil occurs.

At a higher electron energy of 30 MeV as shown in Fig.~\ref{fig:5}(g), a significant portion of photons being emitted at angles exceeds $90^\circ$ with RR, while no photon emitted exceeds $90^\circ$ without RR. This provides a clear experimental verifiable signal for RR below the CRDR, that can be easily realized by placing a single-photon CCD at a fixed angle. This is consistent with the electron trajectory mentioned above and the conclusion in Fig.~\ref{fig:4}. 

As shown in Fig.~\ref{fig:5}(h), the value of $E_c$ for the LL model is higher than that for the MLL model, and this difference becomes increasingly pronounced as the energy increases. The ratio of final to initial total electron energy after collision (black lines) is shown in the black lines. When the ratio is greater than 1, it means that the electron energy gain from the laser is greater than the loss due to RR. The ratio for the MLL model is higher, which also indicates that the quantum corrections suppress the radiation power.
Experimentally verifying the systematic overestimation of radiated power and energy loss by the classical RR, provides a critical test for the validity of quantum correction and the Gaunt factor $g(\chi_e)$.
Furthermore, the decrease in the ratio with increasing energy of the incident electron also indicates that the RR-induced energy loss has begun to dominate.
Experimentally measuring the radiation from electrons with different initial energies can test the relativistic electron dynamics, as well as provides the evidence for RR.

\begin{figure*}
    \centering
    \includegraphics[width=0.75\textwidth]{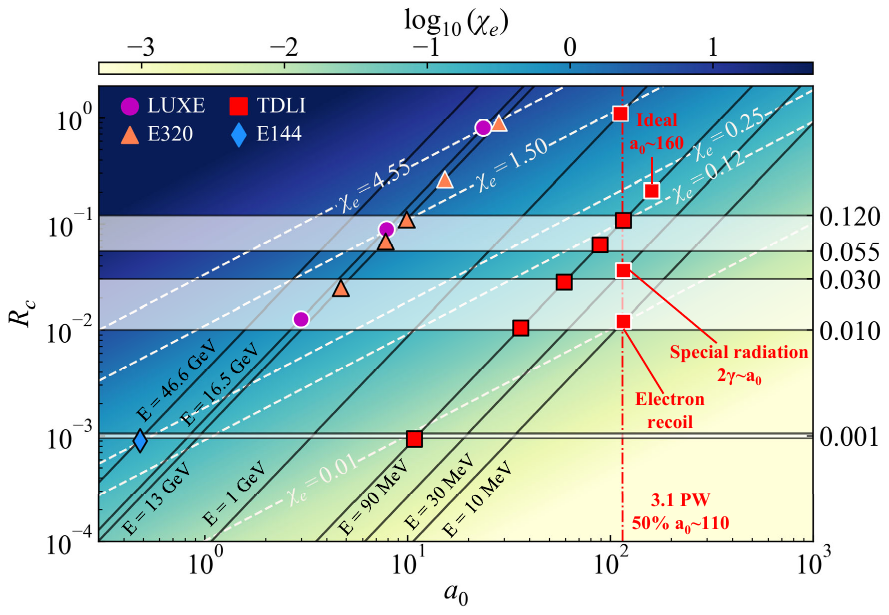}
    \caption{\label{fig:6} Phase-space of $a_0$, $R_c$ and $\chi_e$ in the laser-linac experimental setup around the world. Points with black borders denote existing experimental conditions or completed experiments, whereas points with white borders denote those requiring upgrades. }
\end{figure*}

In our proposed RR experiment within the CRDR, the interaction between PW lasers and tens-of-MeV electron beams enables the observation of strong classical RR signatures and access to the classical-quantum transition. In particular, the unique parameter regime with $R_c \sim 0.1$ and $\chi_e \sim 0.1$ for the RR experiment has never been reported before.
Thus, the upcoming experiment mentioned in this paper occupies a unique position in the experimental parameter space around the world, as shown in Fig.~\ref{fig:6}.
The figure maps the key parameters $a_0$, $R_c$ and $\chi_e$ of laser-linac experimental setup worldwide (black borders: operational; white borders: planned or upgrading). It can be seen that the configuration of high intensity laser and moderate-energy electron beam in TDLI will prominently highlight the significance of the invariant classical RR parameter $R_c$ (as $R_c \propto a_0^2$). This configuration decouples the classical and quantum aspects of RR.
Compared with facilities such as LUXE and E-320, which employ high-energy electron beams to achieve high $\chi_e$ and enter the SFQED-dominated regime, TDLI is engineered to reach the same $R_c$ but a much smaller $\chi_e$, providing a unique aspect to study strong RR effects that are more aligned with the classical electrodynamics framework. It also serves as a bridge between pure classical electrodynamics and the full quantum electrodynamics for testing semi-classical RR theories.
Furthermore, a planned experimental scheme based on injecting linac electrons into a laser wakefield accelerator \cite{Wu2021NP} is commissioned with aims to generate high-quality 1 GeV electron beams. 
This future upgrade will extend the capacity to investigate laser-particle interaction within the SFQED regime ($\chi_e > 1$), such as quantum RR\cite{Dinu2016PRL, Niel2018PRE}, nonlinear Compton scattering\cite{King2020PRA, Li2015PRL}, and pair production\cite{Fedotov2010PRL, Breit1934PR,zhuDenseGeVElectron2016}.

\section{conclusion}
We present a novel and currently unique experimental plan for observing the RR effects within the CRDR. The combination of a PW-laser and a tens-of-MeV electron bunch yields key parameters $R_c \sim 0.1$ and $\chi_e \sim 0.1$. 
This unique parameter space exhibiting strong RR with weak quantum effect, enables the separation of the classical RR effects from other SFQED processes, and provides the access to the classical-quantum transition.
In addition, 3D-PIC simulations show that the interaction intensity can be tuned via adjusting the time delay for the collision or altering the electron energy. The tunable interaction intensity provides distinct signals for RR, such as changes in charge quantity or radiation at specific angles, which are detectable experimentally. Our findings pave the way for precision tests of radiation reaction across the classical-quantum transition, offering critical insights to deepen our understanding of strong-field quantum electrodynamics.

\begin{acknowledgments}
We thank Dao Xiang from Shanghai Jiao Tong University for fruitful discussion. This work is supported by the National Key Programme for S\&T Research and Development of China (2021YFA1601700), the National Natural Science Foundation of China (Grant No.12135009, 12375244), the Natural Science Foundation of Hunan Province of China (Grant No.2025JJ30002), Fundamental and Interdisciplinary Disciplines Breakthrough Plan of the Ministry of Education of China (JYB2025XDXM204) and AI for Science Program of Shanghai Municipal Commission of Economy and Informatization (Grand No.2025-GZL-RGZN-BTBX-02029). We thank the sponsorship from Yangyang Development Fund.

\end{acknowledgments}

\bibliography{main}

@misc{eli,
    title     = "{The Extreme Light Infrastructure. https://www.eli-laser.eu}",
    url       = "https://www.eli-laser.eu"
}

@book{Abraham1905,
    author = "M. Abraham",
    title = "Theorie der Elektrizit{\"a}t, Vol. II",
    publisher = "Teubner, Leipzig",
    year = "1905"
}

@book{Lorentz1909,
    author = "H. A. Lorentz",
    title = "The Theory of Electrons",
    publisher = "Teubner, Leipzig",
    year = "1909"
}

@article{Breit1934PR,
  title = {Collision of Two Light Quanta},
  author = {Breit, G. and Wheeler, John A.},
  journal = {Phys. Rev.},
  volume = {46},
  issue = {12},
  pages = {1087--1091},
  numpages = {0},
  year = {1934},
  month = {Dec},
  publisher = {American Physical Society},
  doi = {10.1103/PhysRev.46.1087},
  url = {https://link.aps.org/doi/10.1103/PhysRev.46.1087}
}

@article{Dirac1938,
    author = "P. A. M. Dirac",
    title = "Classical theory of radiating electrons",
    journal = "Proc. R. Soc. A",
    year = "1938",
    volume  = "167",
    number  = "929",
    pages   = "148-169"
}

@article{PhysRev.82.664,
  title = {On Gauge Invariance and Vacuum Polarization},
  author = {Schwinger, Julian},
  journal = {Phys. Rev.},
  volume = {82},
  issue = {5},
  pages = {664--679},
  numpages = {0},
  year = {1951},
  month = {Jun},
  publisher = {American Physical Society},
  doi = {10.1103/PhysRev.82.664},
  url = {https://link.aps.org/doi/10.1103/PhysRev.82.664}
}

@article{ErberRevModPhys.38.626,
  title = {High-Energy Electromagnetic Conversion Processes in Intense Magnetic Fields},
  author = {Erber, Thomas},
  journal = {Rev. Mod. Phys.},
  volume = {38},
  issue = {4},
  pages = {626--659},
  numpages = {0},
  year = {1966},
  month = {Oct},
  publisher = {American Physical Society},
  doi = {10.1103/RevModPhys.38.626}
}

@book{Landau1975,
    author = "Landau, L. D. and Lifshitz, E. M.",
    title = "The Classical Theory of Fields",
    publisher = "Pergamon",
    year = "1975"
}

@article{Tajima1979PRL,
  title = {Laser Electron Accelerator},
  author = {Tajima, T. and Dawson, J. M.},
  journal = {Phys. Rev. Lett.},
  volume = {43},
  issue = {4},
  pages = {267--270},
  numpages = {0},
  year = {1979},
  month = {Jul},
  publisher = {American Physical Society},
  doi = {10.1103/PhysRevLett.43.267},
  url = {https://link.aps.org/doi/10.1103/PhysRevLett.43.267}
}

@article{Ritus.jrlr.1985,
  title = {Quantum effects of the interaction of elementary particles with an intense electromagnetic field},
  author = {Ritus, V. I.},
  journal = {J. Russ. Laser Res.},
  volume = {6},
  issue = {5},
  pages = {497--617},
  year = {1985},
  month = {Sept},
  doi = {10.1007/BF01120220}
}

@article{STRICKLAND1985219,
title = {Compression of amplified chirped optical pulses},
journal = {Optics Communications},
volume = {56},
number = {3},
pages = {219-221},
year = {1985},
issn = {0030-4018},
doi = {https://doi.org/10.1016/0030-4018(85)90120-8},
url = {https://www.sciencedirect.com/science/article/pii/0030401885901208},
author = {Donna Strickland and Gerard Mourou},
}

@article{Ford1991PLA,
  title = {Radiation reaction in electrodynamics and the elimination of runaway solutions},
  author = {G. W. Ford and R. F. O'Connell},
  journal = {Physics Letters A},
  volume = {157},
  number = {4},
  pages = {217-220},
  year = {1991},
  issn = {0375-9601},
  doi = {10.1016/0375-9601(91)90054-C}
}

@book{Peskin.qft.1995,
  author		= {Peskin, M. E. and Schroeder, D. V.},
  title		= {An Introduction To Quantum Field Theory},
  address		= {Boulder},
  publisher	= {Westview Press},
  year		= {1995},
  doi         = {10.1201/9780429503559}
}

@article{PhysRevLett.76.3116,
  title = {Observation of Nonlinear Effects in Compton Scattering},
  author = {Bula, C. and McDonald, K. T. and Prebys, E. J. and Bamber, C. and Boege, S. and Kotseroglou, T. and Melissinos, A. C. and Meyerhofer, D. D. and Ragg, W. and Burke, D. L. and Field, R. C. and Horton-Smith, G. and Odian, A. C. and Spencer, J. E. and Walz, D. and Berridge, S. C. and Bugg, W. M. and Shmakov, K. and Weidemann, A. W.},
  journal = {Phys. Rev. Lett.},
  volume = {76},
  issue = {17},
  pages = {3116--3119},
  numpages = {0},
  year = {1996},
  month = {Apr},
  publisher = {American Physical Society},
  doi = {10.1103/PhysRevLett.76.3116},
  url = {https://link.aps.org/doi/10.1103/PhysRevLett.76.3116}
}

@article{PhysRevLett.79.1626,
  title = {Positron Production in Multiphoton Light-by-Light Scattering},
  author = {Burke, D. L. and Field, R. C. and Horton-Smith, G. and Spencer, J. E. and Walz, D. and Berridge, S. C. and Bugg, W. M. and Shmakov, K. and Weidemann, A. W. and Bula, C. and McDonald, K. T. and Prebys, E. J. and Bamber, C. and Boege, S. J. and Koffas, T. and Kotseroglou, T. and Melissinos, A. C. and Meyerhofer, D. D. and Reis, D. A. and Ragg, W.},
  journal = {Phys. Rev. Lett.},
  volume = {79},
  issue = {9},
  pages = {1626--1629},
  numpages = {0},
  year = {1997},
  month = {Sep},
  publisher = {American Physical Society},
  doi = {10.1103/PhysRevLett.79.1626},
  url = {https://link.aps.org/doi/10.1103/PhysRevLett.79.1626}
}

@book{Jackson.1998,
  author		= "Jackson, J. D.",
  title		= "Classical {E}lectrodynamics, 3rd ed.",
  address		= "New York",
  publisher	= "Wiley",
  year		= "1998"
}

@article{PhysRevD.60.092004,
  title = {Studies of nonlinear QED in collisions of 46.6 GeV electrons with intense laser pulses},
  author = {Bamber, C. and Boege, S. J. and Koffas, T. and Kotseroglou, T. and Melissinos, A. C. and Meyerhofer, D. D. and Reis, D. A. and Ragg, W. and Bula, C. and McDonald, K. T. and Prebys, E. J. and Burke, D. L. and Field, R. C. and Horton-Smith, G. and Spencer, J. E. and Walz, D. and Berridge, S. C. and Bugg, W. M. and Shmakov, K. and Weidemann, A. W.},
  journal = {Phys. Rev. D},
  volume = {60},
  issue = {9},
  pages = {092004},
  numpages = {43},
  year = {1999},
  month = {Oct},
  publisher = {American Physical Society},
  doi = {10.1103/PhysRevD.60.092004},
  url = {https://link.aps.org/doi/10.1103/PhysRevD.60.092004}
}

@article{Salamin2002PRL,
  title = {Electron Acceleration by a Tightly Focused Laser Beam},
  author = {Salamin, Yousef I. and Keitel, Christoph H.},
  journal = {Phys. Rev. Lett.},
  volume = {88},
  issue = {9},
  pages = {095005},
  numpages = {4},
  year = {2002},
  month = {Feb},
  publisher = {American Physical Society},
  doi = {10.1103/PhysRevLett.88.095005},
  url = {https://link.aps.org/doi/10.1103/PhysRevLett.88.095005}
}

@article{Danson.nf.2004,
doi = {10.1088/0029-5515/44/12/S15},
year = {2004},
month = {nov},
publisher = {},
volume = {44},
number = {12},
pages = {S239},
author = {Danson, C.N. and Brummitt, P.A. and Clarke, R.J. and J.L. Collier and B. Fell and A.J. Frackiewicz and S. Hancock and S. Hawkes and C. Hernandez-Gomez and P. Holligan and M.H.R. Hutchinson and A. Kidd and W.J. Lester and I.O. Musgrave and D. Neely and D.R. Neville and P.A. Norreys and D.A. Pepler and C.J. Reason and W. Shaikh and T.B. Winstone and R.W.W. Wyatt and B.E. Wyborn},
title = {Vulcan {P}etawatt-an ultra-high-intensity interaction facility},
journal = {Nuclear Fusion}
}

@article{Piazza.2008.lmp,
  title = {Exact Solution of the Landau-Lifshitz Equation in a Plane Wave},
  author = {Di Piazza, A.},
  journal = {Lett. Math. Phys.},
  volume = {83},
  pages = {305--313},
  numpages = {9},
  year = {2008},
  doi = {10.1007/s11005-008-0228-9}
}

@article{Sokolov2009jetp,
  title = {Renormalization of the Lorentz-Abraham-Dirac equation for radiation reaction force in classical electrodynamics},
  author = {Sokolov, I. V.},
  journal = {Journal of Experimental and Theoretical Physics},
  volume = {109},
  pages = {207},
  year = {2009},
  doi = {10.1134/S1063776109080044}
}

@article{Piazza2009PRL,
  title = {Strong Signatures of Radiation Reaction below the Radiation-Dominated Regime},
  author = {Di Piazza, A. and Hatsagortsyan, K. Z. and Keitel, C. H.},
  journal = {Phys. Rev. Lett.},
  volume = {102},
  issue = {25},
  pages = {254802},
  numpages = {4},
  year = {2009},
  month = {Jun},
  publisher = {American Physical Society},
  doi = {10.1103/PhysRevLett.102.254802},
  url = {https://link.aps.org/doi/10.1103/PhysRevLett.102.254802}
}

@article{Jaroschek2009PRL,
  title = {Radiation-Dominated Relativistic Current Sheets},
  author = {Jaroschek, C. H. and Hoshino, M.},
  journal = {Phys. Rev. Lett.},
  volume = {103},
  issue = {7},
  pages = {075002},
  numpages = {4},
  year = {2009},
  month = {Aug},
  publisher = {American Physical Society},
  doi = {10.1103/PhysRevLett.103.075002},
  url = {https://link.aps.org/doi/10.1103/PhysRevLett.103.075002}
}

@article{Kirk2009PPCF,
doi = {10.1088/0741-3335/51/8/085008},
url = {https://doi.org/10.1088/0741-3335/51/8/085008},
year = {2009},
month = {jul},
publisher = {},
volume = {51},
number = {8},
pages = {085008},
author = {Kirk, J G and Bell, A R and Arka, I},
title = {Pair production in counter-propagating laser beams},
journal = {Plasma Physics and Controlled Fusion},
}

@article{vulcan.2010,
  title = {The {Vulcan} 10 {PW} project},
  author = {C. Hernandez-Gomez and S. P. Blake and O. Chekhlov and R. J. Clarke and A. M. Dunne and M. Galimberti and S. Hancock and R. Heathcote and P. Holligan and A. Lyachev and P. Matousek and I. O. Musgrave and D. Neely and P. A. Norreys and I. Ross and Y. Tang and T. B. Winstone and B. E. Wyborn and J. Collier},
  journal = {J. Phys.: Conf. Ser.},
  volume = {244},
  pages = {032006},
  year = {2010},
  doi = {10.1088/1742-6596/244/3/032006}
}

@article{Fedotov2010PRL,
  title = {Limitations on the Attainable Intensity of High Power Lasers},
  author = {Fedotov, A. M. and Narozhny, N. B. and Mourou, G. and Korn, G.},
  journal = {Phys. Rev. Lett.},
  volume = {105},
  issue = {8},
  pages = {080402},
  numpages = {4},
  year = {2010},
  month = {Aug},
  publisher = {American Physical Society},
  doi = {10.1103/PhysRevLett.105.080402},
  url = {https://link.aps.org/doi/10.1103/PhysRevLett.105.080402}
}

@article{PhysRevLett.105.220403,
  title = {Quantum Radiation Reaction Effects in Multiphoton Compton Scattering},
  author = {Di Piazza, A. and Hatsagortsyan, K. Z. and Keitel, C. H.},
  journal = {Phys. Rev. Lett.},
  volume = {105},
  issue = {22},
  pages = {220403},
  numpages = {4},
  year = {2010},
  month = {Nov},
  publisher = {American Physical Society},
  doi = {10.1103/PhysRevLett.105.220403},
  url = {https://link.aps.org/doi/10.1103/PhysRevLett.105.220403}
}

@article{Duclous2011PPCF,
doi = {10.1088/0741-3335/53/1/015009},
url = {https://doi.org/10.1088/0741-3335/53/1/015009},
year = {2010},
month = {nov},
publisher = {},
volume = {53},
number = {1},
pages = {015009},
author = {Duclous, R and Kirk, J G and Bell, A R},
title = {Monte Carlo calculations of pair production in high-intensity laser–plasma interactions},
journal = {Plasma Physics and Controlled Fusion}
}

@article{Tavani2011science,
author = {M. Tavani  and A. Bulgarelli  and V. Vittorini  and A. Pellizzoni  and E. Striani  and P. Caraveo  and M. C. Weisskopf  and A. Tennant  and G. Pucella  and A. Trois  and E. Costa  and Y. Evangelista  and C. Pittori  and F. Verrecchia  and E. Del Monte  and R. Campana  and M. Pilia  and A. De Luca  and I. Donnarumma  and D. Horns  and C. Ferrigno  and C. O. Heinke  and M. Trifoglio  and F. Gianotti  and S. Vercellone  and A. Argan  and G. Barbiellini  and P. W. Cattaneo  and A. W. Chen  and T. Contessi  and F. D’Ammando  and G. DeParis  and G. Di Cocco  and G. Di Persio  and M. Feroci  and A. Ferrari  and M. Galli  and A. Giuliani  and M. Giusti  and C. Labanti  and I. Lapshov  and F. Lazzarotto  and P. Lipari  and F. Longo  and F. Fuschino  and M. Marisaldi  and S. Mereghetti  and E. Morelli  and E. Moretti  and A. Morselli  and L. Pacciani  and F. Perotti  and G. Piano  and P. Picozza  and M. Prest  and M. Rapisarda  and A. Rappoldi  and A. Rubini  and S. Sabatini  and P. Soffitta  and E. Vallazza  and A. Zambra  and D. Zanello  and F. Lucarelli  and P. Santolamazza  and P. Giommi  and L. Salotti  and G. F. Bignami },
title = {Discovery of Powerful Gamma-Ray Flares from the Crab Nebula},
journal = {Science},
volume = {331},
number = {6018},
pages = {736-739},
year = {2011},
doi = {10.1126/science.1200083},
URL = {https://www.science.org/doi/abs/10.1126/science.1200083},
}

@article{Piazza.rmp.2012,
  title = {Extremely high-intensity laser interactions with fundamental quantum systems},
  author = {Di Piazza, A. and M\"uller, C. and Hatsagortsyan, K. Z. and Keitel, C. H.},
  journal = {Rev. Mod. Phys.},
  volume = {84},
  issue = {3},
  pages = {1177--1228},
  numpages = {0},
  year = {2012},
  month = {Aug},
  publisher = {American Physical Society},
  doi = {10.1103/RevModPhys.84.1177}
}

@article{Thomas2012PRX,
  title = {Strong Radiation-Damping Effects in a Gamma-Ray Source Generated by the Interaction of a High-Intensity Laser with a Wakefield-Accelerated Electron Beam},
  author = {Thomas, A. G. R. and Ridgers, C. P. and Bulanov, S. S. and Griffin, B. J. and Mangles, S. P. D.},
  journal = {Phys. Rev. X},
  volume = {2},
  issue = {4},
  pages = {041004},
  numpages = {13},
  year = {2012},
  month = {Oct},
  publisher = {American Physical Society},
  doi = {10.1103/PhysRevX.2.041004},
  url = {https://link.aps.org/doi/10.1103/PhysRevX.2.041004}
}

@article{Corde2013RMP,
  title = {Femtosecond x rays from laser-plasma accelerators},
  author = {Corde, S. and Ta Phuoc, K. and Lambert, G. and Fitour, R. and Malka, V. and Rousse, A. and Beck, A. and Lefebvre, E.},
  journal = {Rev. Mod. Phys.},
  volume = {85},
  issue = {1},
  pages = {1--48},
  numpages = {0},
  year = {2013},
  month = {Jan},
  publisher = {American Physical Society},
  doi = {10.1103/RevModPhys.85.1},
  url = {https://link.aps.org/doi/10.1103/RevModPhys.85.1}
}

@article{Neitz2013PRL,
  title = {Stochasticity Effects in Quantum Radiation Reaction},
  author = {Neitz, N. and Di Piazza, A.},
  journal = {Phys. Rev. Lett.},
  volume = {111},
  issue = {5},
  pages = {054802},
  numpages = {5},
  year = {2013},
  month = {Aug},
  publisher = {American Physical Society},
  doi = {10.1103/PhysRevLett.111.054802},
  url = {https://link.aps.org/doi/10.1103/PhysRevLett.111.054802}
}

@article{Burton2014cp,
  title = {Aspects of electromagnetic radiation reaction in strong fields},
  author = {Burton, D. A. and Noble, A.},
  journal = {Contemporary Physics},
  volume = {55},
  number = {2},
  pages = {110--121},
  year = {2014},
  doi = {10.1080/00107514.2014.886840}
}

@article{Vranic2014PRL,
  title = {All-Optical Radiation Reaction at $1{0}^{21}\text{ }\text{ }\mathrm{W}/{\mathrm{cm}}^{2}$},
  author = {Vranic, M. and Martins, J. L. and Vieira, J. and Fonseca, R. A. and Silva, L. O.},
  journal = {Phys. Rev. Lett.},
  volume = {113},
  issue = {13},
  pages = {134801},
  numpages = {5},
  year = {2014},
  month = {Sep},
  publisher = {American Physical Society},
  doi = {10.1103/PhysRevLett.113.134801},
  url = {https://link.aps.org/doi/10.1103/PhysRevLett.113.134801}
}

@article{RIDGERS2014273,
title = {Modelling gamma-ray photon emission and pair production in high-intensity laser–matter interactions},
journal = {Journal of Computational Physics},
volume = {260},
pages = {273-285},
year = {2014},
issn = {0021-9991},
doi = {https://doi.org/10.1016/j.jcp.2013.12.007},
url = {https://www.sciencedirect.com/science/article/pii/S0021999113008061},
author = {C.P. Ridgers and J.G. Kirk and R. Duclous and T.G. Blackburn and C.S. Brady and K. Bennett and T.D. Arber and A.R. Bell}
}

@article{Blackburn.PhysRevLett.112.015001,
  title = {Quantum Radiation Reaction in Laser--Electron-Beam Collisions},
  author = {Blackburn, T. G. and Ridgers, C. P. and Kirk, J. G. and Bell, A. R.},
  journal = {Phys. Rev. Lett.},
  volume = {112},
  issue = {1},
  pages = {015001},
  numpages = {5},
  year = {2014},
  month = {Jan},
  publisher = {American Physical Society},
  doi = {10.1103/PhysRevLett.112.015001},
  url = {https://link.aps.org/doi/10.1103/PhysRevLett.112.015001}
}

@article{Ji2014PRL,
  title = {Radiation-Reaction Trapping of Electrons in Extreme Laser Fields},
  author = {Ji, L. L. and Pukhov, A. and Kostyukov, I. Yu. and Shen, B. F. and Akli, K.},
  journal = {Phys. Rev. Lett.},
  volume = {112},
  issue = {14},
  pages = {145003},
  numpages = {5},
  year = {2014},
  month = {Apr},
  publisher = {American Physical Society},
  doi = {10.1103/PhysRevLett.112.145003},
  url = {https://link.aps.org/doi/10.1103/PhysRevLett.112.145003}
}

@article{Corvan2014RSI,
    author = {Corvan, D. J. and Sarri, G. and Zepf, M.},
    title = {Design of a compact spectrometer for high-flux MeV gamma-ray beams},
    journal = {Review of Scientific Instruments},
    volume = {85},
    number = {6},
    pages = {065119},
    year = {2014},
    month = {06},
    issn = {0034-6748},
    doi = {10.1063/1.4884643},
    url = {https://doi.org/10.1063/1.4884643},
}

@article{Li2015PRL,
  title = {Attosecond Gamma-Ray Pulses via Nonlinear Compton Scattering in the Radiation-Dominated Regime},
  author = {Li, Jian-Xing and Hatsagortsyan, Karen Z. and Galow, Benjamin J. and Keitel, Christoph H.},
  journal = {Phys. Rev. Lett.},
  volume = {115},
  issue = {20},
  pages = {204801},
  numpages = {5},
  year = {2015},
  month = {Nov},
  publisher = {American Physical Society},
  doi = {10.1103/PhysRevLett.115.204801},
  url = {https://link.aps.org/doi/10.1103/PhysRevLett.115.204801}
}

@article{Arber.ppcf.2025,
  title = {Contemporary particle-in-cell approach to laser-plasma modelling},
  author = {Arber, T. D. and Bennett, K. and Brady, C. S. and Lawrence-Douglas, A. and Ramsay, M. G. and Sircombe, N. J. and Gillies, P. and Evans, R. G. and Schmitz, H. and Bell, A. R. and Ridgers, C. P.},
  journal = {Plasma Physics and Controlled Fusion},
  volume = {57},
  number = {11},
  pages = {113001},
  year = {2015},
  doi = {10.1088/0741-3335/57/11/113001}
}

@article{zhuDenseGeVElectron2016,
  title = {Dense {{GeV}} Electron\textendash Positron Pairs Generated by Lasers in near-Critical-Density Plasmas},
  author = {Zhu, Xing-Long and Yu, Tong-Pu and Sheng, Zheng-Ming and Yin, Yan and Turcu, Ion Cristian Edmond and Pukhov, Alexander},
  year = {2016},
  month = dec,
  journal = {Nat Commun},
  volume = {7},
  number = {1},
  pages = {13686},
  langid = {english}
}

@article{Dinu2016PRL,
  title = {Quantum Radiation Reaction: From Interference to Incoherence},
  author = {Dinu, Victor and Harvey, Chris and Ilderton, Anton and Marklund, Mattias and Torgrimsson, Greger},
  journal = {Phys. Rev. Lett.},
  volume = {116},
  issue = {4},
  pages = {044801},
  numpages = {6},
  year = {2016},
  month = {Jan},
  publisher = {American Physical Society},
  doi = {10.1103/PhysRevLett.116.044801},
  url = {https://link.aps.org/doi/10.1103/PhysRevLett.116.044801}
}

@INPROCEEDINGS{Sung2016IPC,
  author={Sung, Jae Hee and Lee, Seong Ku and Lee, Hwang Woon and Yoo, Je Yoon and Nam, Chang Hee},
  booktitle={2016 IEEE Photonics Conference (IPC)}, 
  title={High-contrast 0.1-Hz 4-PW laser at CoReLS}, 
  year={2016},
  volume={},
  number={},
  pages={456-457},
  doi={10.1109/IPCon.2016.7831176}}

@article{Yan2017NP,
  title = {High-order multiphoton Thomson scattering},
  author = {Yan, Wenchao and Fruhling, Colton and Golovin, Grigory and Haden, Daniel and Luo, Ji and Zhang, Ping and Zhao, Baozhen and Zhang, Jun and Liu, Cheng and Chen, Min and Chen, Shouyuan and Banerjee, Sudeep and Umstadter, Donald},
  journal = {Nature Photonics},
  volume = {11},
  number = {8},
  pages = {514--520},
  year = {2017},
  doi = {10.1038/nphoton.2017.100},
  url = {https://doi.org/10.1038/nphoton.2017.100},
  issn = {1749-4893}
}

@article{Samarin2017JMO,
author = {G. M. Samarin and M. Zepf and G. Sarri},
title = {Radiation reaction studies in an all-optical set-up: experimental limitations},
journal = {Journal of Modern Optics},
volume = {65},
number = {11},
pages = {1362--1369},
year = {2017},
publisher = {Taylor \& Francis},
doi = {10.1080/09500340.2017.1353655},

}

@book{Lecoq2017Inorganic,
  title={Inorganic Scintillators for Detector Systems: Physical Principles and Crystal Engineering},
  author={Lecoq, Paul and Gektin, Alexander and Korzhik, Mikhail},
  edition={2},
  series={Particle Acceleration and Detection},
  publisher={Springer Cham},
  year={2017},
  doi={10.1007/978-3-319-45522-8},
  isbn={978-3-319-45522-8}
}

@article{Behm2018RSI,
    author = {Behm, K. T. and Cole, J. M. and Joglekar, A. S. and Gerstmayr, E. and Wood, J. C. and Baird, C. D. and Blackburn, T. G. and Duff, M. and Harvey, C. and Ilderton, A. and Kuschel, S. and Mangles, S. P. D. and Marklund, M. and McKenna, P. and Murphy, C. D. and Najmudin, Z. and Poder, K. and Ridgers, C. P. and Sarri, G. and Samarin, G. M. and Symes, D. and Warwick, J. and Zepf, M. and Krushelnick, K. and Thomas, A. G. R.},
    title = {A spectrometer for ultrashort gamma-ray pulses with photon energies greater than 10 MeV},
    journal = {Review of Scientific Instruments},
    volume = {89},
    number = {11},
    pages = {113303},
    year = {2018},
    month = {11},
    issn = {0034-6748},
    doi = {10.1063/1.5056248},
    url = {https://doi.org/10.1063/1.5056248},
}

@article{Li2018ol,
  author = {Wenqi Li and Zebiao Gan and Lianghong Yu and Cheng Wang and Yanqi Liu and Zhen Guo and Lu Xu and Min Xu and Yin Hang and Yi Xu and Jianye Wang and Pei Huang and He Cao and Bo Yao and Xiaobo Zhang and Lingru Chen and Yunhai Tang and Shuai Li and Xingyan Liu and Shanming Li and Mingzhu He and Dinjun Yin and Xiaoyan Liang and Yuxin Leng and Ruxin Li and Zhizhan Xu},
  journal = {Opt. Lett.},
  number = {22},
  pages = {5681--5684},
  title = {339J high-energy Ti:sapphire chirped-pulse amplifier for 10PW laser facility},
  volume = {43},
  year = {2018},
  doi = {10.1364/OL.43.005681}
}

@article{Cole2018PRX,
  title = {Experimental Evidence of Radiation Reaction in the Collision of a High-Intensity Laser Pulse with a Laser-Wakefield Accelerated Electron Beam},
  author = {Cole, J. M. and Behm, K. T. and Gerstmayr, E. and Blackburn, T. G. and Wood, J. C. and Baird, C. D. and Duff, M. J. and Harvey, C. and Ilderton, A. and Joglekar, A. S. and Krushelnick, K. and Kuschel, S. and Marklund, M. and McKenna, P. and Murphy, C. D. and Poder, K. and Ridgers, C. P. and Samarin, G. M. and Sarri, G. and Symes, D. R. and Thomas, A. G. R. and Warwick, J. and Zepf, M. and Najmudin, Z. and Mangles, S. P. D.},
  journal = {Phys. Rev. X},
  volume = {8},
  issue = {1},
  pages = {011020},
  numpages = {11},
  year = {2018},
  month = {Feb},
  publisher = {American Physical Society},
  doi = {10.1103/PhysRevX.8.011020},
  url = {https://link.aps.org/doi/10.1103/PhysRevX.8.011020}
}

@article{Poder2018PRX,
  title = {Experimental Signatures of the Quantum Nature of Radiation Reaction in the Field of an Ultraintense Laser},
  author = {Poder, K. and Tamburini, M. and Sarri, G. and Di Piazza, A. and Kuschel, S. and Baird, C. D. and Behm, K. and Bohlen, S. and Cole, J. M. and Corvan, D. J. and Duff, M. and Gerstmayr, E. and Keitel, C. H. and Krushelnick, K. and Mangles, S. P. D. and McKenna, P. and Murphy, C. D. and Najmudin, Z. and Ridgers, C. P. and Samarin, G. M. and Symes, D. R. and Thomas, A. G. R. and Warwick, J. and Zepf, M.},
  journal = {Phys. Rev. X},
  volume = {8},
  issue = {3},
  pages = {031004},
  numpages = {11},
  year = {2018},
  month = {Jul},
  publisher = {American Physical Society},
  doi = {10.1103/PhysRevX.8.031004},
  url = {https://link.aps.org/doi/10.1103/PhysRevX.8.031004}
}

@article{Niel2018PRE,
  title = {From quantum to classical modeling of radiation reaction: A focus on stochasticity effects},
  author = {Niel, F. and Riconda, C. and Amiranoff, F. and Duclous, R. and Grech, M.},
  journal = {Phys. Rev. E},
  volume = {97},
  issue = {4},
  pages = {043209},
  numpages = {27},
  year = {2018},
  month = {Apr},
  publisher = {American Physical Society},
  doi = {10.1103/PhysRevE.97.043209},
  url = {https://link.aps.org/doi/10.1103/PhysRevE.97.043209}
}

@article{dansonPetawattExawattClass2019a,
  title = {Petawatt and Exawatt Class Lasers Worldwide},
  author = {Danson, Colin N. and Haefner, Constantin and Bromage, Jake and Butcher, Thomas and Chanteloup, Jean-Christophe F. and Chowdhury, Enam A. and Galvanauskas, Almantas and Gizzi, Leonida A. and Hein, Joachim and Hillier, David I. and Hopps, Nicholas W. and Kato, Yoshiaki and Khazanov, Efim A. and Kodama, Ryosuke and Korn, Georg and Li, Ruxin and Li, Yutong and Limpert, Jens and Ma, Jingui and Nam, Chang Hee and Neely, David and Papadopoulos, Dimitrios and Penman, Rory R. and Qian, Liejia and Rocca, Jorge J. and Shaykin, Andrey A. and Siders, Craig W. and Spindloe, Christopher and Szatm{\'a}ri, S{\'a}ndor and Trines, Raoul M. G. M. and Zhu, Jianqiang and Zhu, Ping and Zuegel, Jonathan D.},
  journal = {High Pow. Laser Sci. Eng.},
  volume = {7},
  pages = {e54},
  year = {2019},
  doi = {10.1017/hpl.2019.36}
}

@article{PhysRevLett.122.190404,
  title = {Prospect of Studying Nonperturbative {QED} with Beam-Beam Collisions},
  author = {Yakimenko, V. and Meuren, S. and Del Gaudio, F. and Baumann, C. and Fedotov, A. and Fiuza, F. and Grismayer, T. and Hogan, M. J. and Pukhov, A. and Silva, L. O. and White, G.},
  journal = {Phys. Rev. Lett.},
  volume = {122},
  issue = {19},
  pages = {190404},
  numpages = {7},
  year = {2019},
  month = {May},
  publisher = {American Physical Society},
  doi = {10.1103/PhysRevLett.122.190404},
  url = {https://link.aps.org/doi/10.1103/PhysRevLett.122.190404}
}

@article{PhysRevAccelBeams.22.101301,
  title = {{FACET-II} facility for advanced accelerator experimental tests},
  author = {Yakimenko, V. and Alsberg, L. and Bong, E. and Bouchard, G. and Clarke, C. and Emma, C. and Green, S. and Hast, C. and Hogan, M. J. and Seabury, J. and Lipkowitz, N. and O'Shea, B. and Storey, D. and White, G. and Yocky, G.},
  journal = {Phys. Rev. Accel. Beams},
  volume = {22},
  issue = {10},
  pages = {101301},
  numpages = {11},
  year = {2019},
  month = {Oct},
  publisher = {American Physical Society},
  doi = {10.1103/PhysRevAccelBeams.22.101301},
  url = {https://link.aps.org/doi/10.1103/PhysRevAccelBeams.22.101301}
}

@article{Danson2019HPLSE,
    title={Petawatt and exawatt class lasers worldwide},
    author={Danson, Colin N. and Haefner, Constantin and Bromage, Jake and Butcher, Thomas and Chanteloup, Jean-Christophe F. and Chowdhury, Enam A. and Galvanauskas, Almantas and Gizzi, Leonida A. and Hein, Joachim and Hillier, David I. and others},
    journal={High Power Laser Science and Engineering},
    volume={7},
    pages={e54},
    year={2019},
    doi={10.1017/hpl.2019.36}
}

@article{Geng2019CP,
  author    = {Geng, X. S. and Ji, L. L. and Shen, B. F. and Feng, B. and Guo, Z. and Yu, Q. and Zhang, L. G. and Xu, Z. Z.},
  title     = {Quantum reflection above the classical radiation-reaction barrier in the quantum electro-dynamics regime},
  journal   = {Communications Physics},
  year      = {2019},
  date      = {2019/06/21},
  volume    = {2},
  number    = {1},
  pages     = {66},
  issn      = {2399-3650},
  doi       = {10.1038/s42005-019-0164-2},
  url       = {https://doi.org/10.1038/s42005-019-0164-2},
}

@article{Schwinkendorf_2019,
doi = {10.1088/1748-0221/14/09/P09025},
url = {https://doi.org/10.1088/1748-0221/14/09/P09025},
year = {2019},
month = {sep},
publisher = {},
volume = {14},
number = {09},
pages = {P09025},
author = {Schwinkendorf, J.-P. and Bohlen, S. and Couperus Cabadağ, J.P. and Ding, H. and Irman, A. and Karsch, S. and Köhler, A. and Krämer, J.M. and Kurz, T. and Kuschel, S. and Osterhoff, J. and Schaper, L.F. and Schinkel, D. and Schramm, U. and Zarini, O. and D'Arcy, R.},
title = {Charge calibration of DRZ scintillation phosphor screens},
journal = {Journal of Instrumentation},
}

@article{King2020PRA,
  title = {Nonlinear Compton scattering of polarized photons in plane-wave backgrounds},
  author = {King, B. and Tang, S.},
  journal = {Phys. Rev. A},
  volume = {102},
  issue = {2},
  pages = {022809},
  numpages = {12},
  year = {2020},
  month = {Aug},
  publisher = {American Physical Society},
  doi = {10.1103/PhysRevA.102.022809},
  url = {https://link.aps.org/doi/10.1103/PhysRevA.102.022809}
}

@article{Hu2020PRA,
  title = {Quantum-stochasticity-induced asymmetry in the angular distribution of electrons in a quasiclassical regime},
  author = {Hu, Guang and Sun, Wei-Qiang and Li, Bing-Jun and Li, Yan-Fei and Wang, Wei-Min and Zhu, Meng and Hu, Hua-Si and Li, Yu-Tong},
  journal = {Phys. Rev. A},
  volume = {102},
  issue = {4},
  pages = {042218},
  numpages = {9},
  year = {2020},
  month = {Oct},
  publisher = {American Physical Society},
  doi = {10.1103/PhysRevA.102.042218},
  url = {https://link.aps.org/doi/10.1103/PhysRevA.102.042218}
}

@article{blackburn2020radiation,
  title = {Radiation reaction in electron-beam interactions with high-intensity lasers},
  author = {T. G. Blackburn},
  journal = {Reviews of Modern Plasma Physics},
  volume = {4},
  number = {1},
  pages = {5},
  year = {2020},
  doi = {10.1007/s41614-020-0042-0}
}

@article{Blackburn.rmpp.2020,
  title = {Radiation reaction in electron–beam interactions with high-intensity lasers},
  author = {Blackburn, T. G.},
  journal = {Rev. Mod. Plasma Phys.},
  volume = {4},
  pages = {5},
  year = {2020},
  doi = {10.1007/s41614-020-0042-0}
}

@article{Liu.epjd.2020,
doi = {10.1140/epjd/e2019-100437-4},
year = {2020},
volume = {74},
pages = {7},
author = {Liu, Ke and Yu, Tong Pu and Zou, De Bin and Xu, Xin Rong and Yin, Yan and Shao, Fu Qiu},
title = {Twisted radiation from nonlinear {T}homson scattering with arbitrary incident angle},
journal = {Eur. Phys. J. D}
}

@article{zhu2020sciadv,
author = {Xing-Long Zhu  and Min Chen  and Su-Ming Weng  and Tong-Pu Yu  and Wei-Min Wang  and Feng He  and Zheng-Ming Sheng  and Paul McKenna  and Dino A. Jaroszynski  and Jie Zhang },
title = {Extremely brilliant GeV $\gamma$-rays from a two-stage laser-plasma accelerator},
journal = {Science Advances},
volume = {6},
number = {22},
pages = {eaaz7240},
year = {2020},
doi = {10.1126/sciadv.aaz7240},
URL = {https://www.science.org/doi/abs/10.1126/sciadv.aaz7240},
eprint = {https://www.science.org/doi/pdf/10.1126/sciadv.aaz7240},
}

@article{Abramowicz2021,
  title = {Conceptual design report for the LUXE experiment},
  author = {Abramowicz, H. and Acosta, U. and Altarelli, M. and Aßmann, R. and Bai, Z. and Behnke, T. and Benhammou, Y. and Blackburn, T. and Boogert, S. and Borysov, O. and Borysova, M. and Brinkmann, R. and Bruschi, M. and Burkart, F. and Büßer, K. and Cavanagh, N. and Davidi, O. and Decking, W. and Dosselli, U. and Elkina, N. and Fedotov, A. and Firlej, M. and Fiutowski, T. and Fleck, K. and Gostkin, M. and Grojean, C. and Hallford, J. and Harsh, H. and Hartin, A. and Heinemann, B. and Heinzl, T. and Helary, L. and Hoffmann, M. and Huang, S. and Huang, X. and Idzik, M. and Ilderton, A. and Jacobs, R. and Kämpfer, B. and King, B. and Lahno, H. and Levanon, A. and Levy, A. and Levy, I. and List, J. and Lohmann, W. and Ma, T. and Macleod, A. J. and Malka, V. and Meloni, F. and Mironov, A. and Morandin, M. and Moron, J. and Negodin, E. and Perez, G. and Pomerantz, I. and Pöschl, R. and Prasad, R. and Quéré, F. and Ringwald, A. and Rödel, C. and Rykovanov, S. and Salgado, F. and Santra, A. and Sarri, G. and Sävert, A. and Sbrizzi, A. and Schmitt, S. and Schramm, U. and Schuwalow, S. and Seipt, D. and Shaimerdenova, L. and Shchedrolosiev, M. and Skakunov, M. and Soreq, Y. and Streeter, M. and Swientek, K. and Tal Hod, N. and Tang, S. and Teter, T. and Thoden, D. and Titov, A. I. and Tolbanov, O. and Torgrimsson, G. and Tyazhev, A. and Wing, M. and Zanetti, M. and Zarubin, A. and Zeil, K. and Zepf, M. and Zhemchukov, A.},
  journal = {The European Physical Journal Special Topics},
  volume = {230},
  number = {11},
  pages = {2445--2560},
  year = {2021},
  date = {2021-10-01},
  doi = {10.1140/epjs/s11734-021-00249-z},
  url = {https://doi.org/10.1140/epjs/s11734-021-00249-z},
  issn = {1951-6401},
}

@article{Hong.mre.2021,
  title = {Commissioning experiment of the high-contrast {SILEX-II} multi-petawatt laser facility},
  author = {Hong, Wei and He, Shukai and Teng, Jian and Deng, Zhigang and Zhang, Zhimeng and Lu, Feng and Zhang, Bo and Zhu, Bin and Dai, Zenghai and Cui, Bo and Wu, Yuchi and Liu, Dongxiao and Qi, Wei and Jiao, Jinlong and Zhang, Faqiang and Yang, Zuhua and Zhang, Feng and Bi, Bi and Zeng, Xiaoming and Zhou, Kainan and Zuo, Yanlei and Huang, Xiaojun and Xie, Na and Guo, Yi and Su, Jingqin and Han, Dan and Mao, Ying and Cao, Leifeng and Zhou, Weimin and Gu, Yuqiu and Jing, Feng and Zhang, Baohan and Cai, Hongbo and He, Minqing and Zheng, Wudi and Zhu, Shaoping and Ma, Wenjun and Wang, Dahui and Shou, Yinren and Yan, Xueqing and Qiao, Bin and Zhang, Yi and Zhong, Congling and Yuan, Xiaohui and Wei, Wenqing },
  journal = {Matter and Radiation at Extremes},
  volume = {6},
  pages = {064401},
  year = {2021},
  doi = {10.1063/5.0016019}
}

@article{Wu2021NP,
author = {Wu, Yipeng and Hua, Jianfei and Zhou, Zheng and Zhang, Jie and Liu, Shuang and Peng, Bo and Fang, Yu and Ning, Xiaonan and Nie, Zan and Li, Fei and Zhang, Chaojie and Pai, Chih-Hao and Du, Yingchao and Lu, Wei and Mori, Warren B. and Joshi, Chan},
journal = {Nature Physics},
title = {High-throughput injection–acceleration of electron bunches from a linear accelerator to a laser wakefield accelerator},
year = {2021},
month = jul,
volume = {17},
number = {7},
pages = {801--806},
issn = {1745-2481},
doi = {10.1038/s41567-021-01202-6},
url = {https://doi.org/10.1038/s41567-021-01202-6},
}

@article{Hannasch2021SR,
  author = {Hannasch, A. and Laso Garcia, A. and LaBerge, M. and Zgadzaj, R. and Köhler, A. and Couperus Cabadağ, J. P. and Zarini, O. and Kurz, T. and Ferrari, A. and Molodtsova, M. and Naumann, L. and Cowan, T. E. and Schramm, U. and Irman, A. and Downer, M. C.},
  title = {Compact spectroscopy of keV to MeV X-rays from a laser wakefield accelerator},
  journal = {Scientific Reports},
  year = {2021},
  volume = {11},
  number = {1},
  pages = {14368},
  date = {2021/07/13},
  issn = {2045-2322},
  doi = {10.1038/s41598-021-93689-5},
  url = {https://doi.org/10.1038/s41598-021-93689-5},
}

@article{Chen2022,
  author = {Chen, Jing and Chen, Ji-Yuan and Chen, Jun-Feng and Chen, Xiang and Fu, Chang-Bo and Guo, Jun and He, Le and He, Zheng-Ting and Khaw, Kim Siang and Li, Jia-Lin and Li, Liang and Li, Shu and Lv, Meng and Liu, Dan-Ning and Liu, Han-Qing and Liu, Kun and Liu, Qi-Bin and Liu, Yang and Lu, Ze-Jia and Mo, Cen and Song, Si-Yuan and Wang, Xiao-Long and Wang, Yu-Feng and Wang, Zhen and Wang, Zi-Rui and Wu, Wei-Hao and Xiang, Dao and Yang, Hai-Jun and Zhang, Jun-Hua and Zhang, Yu-Lei and Zhao, Zhi-Yu and Zhu, Xu-Liang and Zhu, Chun-Xiang and Zhu, Yi-Fan},
  year = {2022},
  month = {11},
  day = {29},
  title = {Prospective study of light dark matter search with a newly proposed DarkSHINE experiment},
  journal = {Science China Physics, Mechanics \& Astronomy},
  pages = {211062},
  volume = {66},
  issue = {1},
  issn = {1869-1927},
  doi = {10.1007/s11433-022-1983-8}
}

@article{petri2022AA,
  title={Particle acceleration and radiation reaction in a strongly magnetised rotating dipole},
  author={P{\'e}tri, J},
  journal={Astronomy \& Astrophysics},
  volume={666},
  pages={A5},
  year={2022}
}

@article{Gonoskov.rmp.2022,
  title = {Charged particle motion and radiation in strong electromagnetic fields},
  author = {Gonoskov, A. and Blackburn, T. G. and Marklund, M. and Bulanov, S. S.},
  journal = {Rev. Mod. Phys.},
  volume = {94},
  issue = {4},
  pages = {045001},
  numpages = {63},
  year = {2022},
  month = {Oct},
  publisher = {American Physical Society},
  doi = {10.1103/RevModPhys.94.045001}
}

@inproceedings{Chen:22,
author = {Zhijiang Chen and Sebastian Meuren and Elias Gerstmayr and Vitaly Yakimenko and Philip H. Bucksbaum and David A. Reis and},
journal = {Optica High-brightness Sources and Light-driven Interactions Congress 2022},
pages = {HF4B.6},
publisher = {Optica Publishing Group},
title = {Preparation of Strong-field {QED} Experiments at {FACET-II}},
year = {2022},
doi = {10.1364/HILAS.2022.HF4B.6}
}

@article{Golovanov2022NJP,
doi = {10.1088/1367-2630/ac53b9},
url = {https://doi.org/10.1088/1367-2630/ac53b9},
year = {2022},
month = {mar},
publisher = {IOP Publishing},
volume = {24},
number = {3},
pages = {033011},
author = {Golovanov, A A and Nerush, E N and Kostyukov, I Yu},
title = {Radiation reaction-dominated regime of wakefield acceleration},
journal = {New Journal of Physics},
}

@misc{MZengponderomotive_scattergithub2022,
   author = {Ming Zeng},
   year = {2022},
   note = {the git repository for a tracking code of particles scattered by single or multiple lasers, with both first-principle and ponderomotive force algorithms},
   howpublished = {\url{https://github.com/mingzeng8/ponderomotive_scatter}}
}

@article{Li2023LPR,
author = {Li, Zhaoyang and Leng, Yuxin and Li, Ruxin},
title = {Further Development of the Short-Pulse Petawatt Laser: Trends, Technologies, and Bottlenecks},
journal = {Laser \& Photonics Reviews},
volume = {17},
number = {1},
pages = {2100705},
doi = {https://doi.org/10.1002/lpor.202100705},
url = {https://onlinelibrary.wiley.com/doi/abs/10.1002/lpor.202100705},
year = {2023}
}

@article{Yu2024rmpp,
  title = {Bright X/$\gamma$-ray emission and lepton pair production by strong laser fields: a review},
  author = {Yu, T. P. and Liu, K. and Zhao, J. and Zhu, X. L. and Lu, Y. and Cao, Y. and Zhang, H. and Shao, F. Q. and Sheng, Z. M.},
  journal = {Reviews of Modern Plasma Physics},
  volume = {8},
  number = {1},
  pages = {24},
  year = {2024},
  doi = {10.1007/s41614-024-00158-3}
}

@misc{los.arXiv2407.12071,
      title={Observation of quantum effects on radiation reaction in strong fields}, 
      author={E. E. Los and E. Gerstmayr and C. Arran and M. J. V. Streeter and C. Colgan and C. C. Cobo and B. Kettle and T. G. Blackburn and N. Bourgeois and L. Calvin and J. Carderelli and N. Cavanagh and S. J. D. Dann A. Di Piazza and R. Fitzgarrald and A. Ilderton and C. H. Keitel and M. Marklund and P. McKenna and C. D. Murphy and Z. Najmudin and P. Parsons and P. P. Rajeev and D. R. Symes and M. Tamburini and A. G. R. Thomas and J. C. Wood and M. Zepf and G. Sarri and C. P. Ridgers and S. P. D Mangles},
      year={2024},
      eprint={2407.12071},
      archivePrefix={arXiv},
      primaryClass={hep-ph},
      url={https://arxiv.org/abs/2407.12071}, 
}

@article{Mirzaie2024NP,
  title = {All-optical nonlinear Compton scattering performed with a multi-petawatt laser},
  author = {Mirzaie, Mohammad and Hojbota, Calin Ioan and Kim, Do Yeon and Pathak, Vishwa Bandhu and Pak, Tae Gyu and Kim, Chul Min and Lee, Hwang Woon and Yoon, Jin Woo and Lee, Seong Ku and Rhee, Yong Joo and Vranic, Marija and Amaro, Oscar and Kim, Ki Yong and Sung, Jae Hee and Nam, Chang Hee},
  journal = {Nature Photonics},
  volume = {18},
  number = {11},
  pages = {1212--1217},
  year = {2024},
  doi = {10.1038/s41566-024-01550-8},
  url = {https://doi.org/10.1038/s41566-024-01550-8},
  issn = {1749-4893}
}

@article{Picksley2024PRL,
  title = {Matched Guiding and Controlled Injection in Dark-Current-Free, 10-GeV-Class, Channel-Guided Laser-Plasma Accelerators},
  author = {Picksley, A. and Stackhouse, J. and Benedetti, C. and Nakamura, K. and Tsai, H. E. and Li, R. and Miao, B. and Shrock, J. E. and Rockafellow, E. and Milchberg, H. M. and Schroeder, C. B. and van Tilborg, J. and Esarey, E. and Geddes, C. G. R. and Gonsalves, A. J.},
  journal = {Phys. Rev. Lett.},
  volume = {133},
  issue = {25},
  pages = {255001},
  numpages = {8},
  year = {2024},
  month = {Dec},
  publisher = {American Physical Society},
  doi = {10.1103/PhysRevLett.133.255001},
  url = {https://link.aps.org/doi/10.1103/PhysRevLett.133.255001}
}

@article{Du2024NP,
author = {Du, Xinyuan and Zhao, Shan and Wang, Lu and Wu, Haodi and Ye, Fan and Xue, Kan-Hao and Peng, Shaoqian and Xia, Jianlong and Sang, Ziru and Zhang, Dongdong and Xiong, Zuping and Zheng, Zhiping and Xu, Ling and Niu, Guangda and Tang, Jiang},
title = {Efficient and ultrafast organic scintillators by hot exciton manipulation},
journal = {Nature Photonics},
year = {2024},
month = {02},
day = {01},
volume = {18},
number = {2},
pages = {162--169},
issn = {1749-4893},
doi = {10.1038/s41566-023-01358-y},
url = {https://doi.org/10.1038/s41566-023-01358-y}
}

@misc{abramowicz2025inputesppuluxeexperiment,
      title={Input to the {ESPPU}: The {LUXE} Experiment}, 
      author={{LUXE Collaboration}},
      year={2025},
      eprint={2504.00873},
      archivePrefix={arXiv},
      primaryClass={hep-ex},
      url={https://arxiv.org/abs/2504.00873}, 
}

@article{Li.HPL.2025,
  title = {High-intensity lasers and research activities in China},
  author = {Li, Yutong and Chen, Liming and Chen, Min and Liu, Feng and Gu, Yuqiu and Guo, Bing and Hua, Jianfei and Huang, Taiwu and Leng, Yuxin and Li, Fei and et al.},
  journal = {High Power Laser Science and Engineering},
  volume = {13},
  pages = {e12},
  year = {2025},
  doi = {10.1017/hpl.2024.69}
}

@article{Chen2025HPLSE,
  title={A platform for all-optical Thomson/Compton scattering with versatile parameters},
  volume={13},
  DOI={10.1017/hpl.2025.36},
  journal={High Power Laser Science and Engineering},
  author={Chen, Siyu and Yan, Wenchao and Zhu, Mingyang and Li, Yaojun and Hu, Xichen and Xu, Hao and Zhou, Weijun and Lu, Guangwei and Wei, Mingxuan and Lu, Lin and et al.},
  year={2025},
  pages={e56}
}

@article{Winkler2025Nature,
  author = {Winkler, P. and Trunk, M. and Hübner, L. and Martinez de la Ossa, A. and Jalas, S. and Kirchen, M. and Agapov, I. and Antipov, S. A. and Brinkmann, R. and Eichner, T. and Ferran Pousa, A. and Hülsenbusch, T. and Palmer, G. and Schnepp, M. and Schubert, K. and Thévenet, M. and Walker, P. A. and Werle, C. and Leemans, W. P. and Maier, A. R.},
  title = {Active energy compression of a laser-plasma electron beam},
  journal = {Nature},
  year = {2025},
  date = {2025/04/01},
  volume = {640},
  number = {8060},
  pages = {907--910},
  issn = {1476-4687},
  url = {https://doi.org/10.1038/s41586-025-08772-y},
  doi = {10.1038/s41586-025-08772-y}
}

\end{document}